\documentclass[aps,prd,superscriptaddress,amsmath,amssymb,amsthm,nofootinbib,preprintnumbers]{revtex4}
\usepackage{amsmath}
\usepackage{graphicx}
\usepackage{epstopdf}
\usepackage{float}
\usepackage{hyperref}
\usepackage{subfigure}
\usepackage{color}
\usepackage[T1]{fontenc}
\usepackage[utf8]{inputenc}
\usepackage[toc,page]{appendix}
\usepackage[usenames,dvipsnames]{xcolor}
\usepackage[normalem]{ulem}
\usepackage{amssymb}
\usepackage{mathrsfs}
\usepackage{grffile}

\begin{document}

\title{Reentrant phase transition in holographic thermodynamics of \\ Born-Infeld AdS black hole}

\preprint{CTP-SCU/2022017}

\author{Ning-Chen Bai}
\email{bainingchen@stu.scu.edu.cn}
\author{Li Song}
\email{songli1984@scu.edu.cn}
\author{Jun Tao}
\email{taojun@scu.edu.cn}

\affiliation{College of Physics, Sichuan University, Chengdu, 610065, China}

%\pacs{bla}

%\date{\today}

\begin{abstract}
In this paper, we study the phase tranisions in the  extended holographic thermodynamics of Born-Infeld AdS black holes. The extended gravitational thermodynamics of  black holes is related to the thermodynamics of the dual field theory by varying both cosmological constant and Newton's constant. With  topological analysis and numerical calculation, we find that the reentrant large-small-large black hole phase transition  still exists in $D=4$. Interestingly, a similar reentrant phase transition is also observed in the dual field theory, between the high-, low- and high-entropy thermal states. For higher dimension $D>4$, only the van der Waals like phase transition is observed, both in the bulk and the dual field theory. 

\end{abstract}

\maketitle
%\newpage 

\tableofcontents

\section{Introduction}\label{intro}

Black hole thermodynamics remains one of the most fascinating areas in the study of gravitational theory. It is particularly interesting to investigate the phase transitions in asymptotically anti–de Sitter (AdS) black holes, motivated by the straightforward definition of thermodynamic equilibrium and its possible interpretation in the context of AdS/CFT correspondence \cite{Maldacena:1997re,Gubser:1998bc,Witten:1998qj}. The Hawking-Page phase transition found between thermal radiation and large AdS black hole \cite{Hawking:1982dh} is the first example, which can be interpreted as the confinement/deconfinement phase
transition of gauge fields \cite{Witten:1998zw}. After that, a van der Waals (VdW) like phase transition between the charged small and large AdS black holes was also uncovered \cite{Chamblin:1999tk,Chamblin:1999hg}. 

An intense topic of research in recent years is the identification of the cosmological constant $\Lambda$ as the thermodynamic pressure $P=-\Lambda/8 \pi G$ \cite{Caldarelli:1999xj,Kastor:2009wy,Dolan:2010ha}. This contributes an additional $V \delta P$ term in the first law of AdS black holes, where $V$ can be interpreted as the thermodynamic volume of the black hole. Known as the extended phase space, it has shed new insights on the thermodynamics and phase transitions of AdS black holes, such as understanding the Hawking–Page phase transition as a solid/liquid transition \cite{Kubiznak:2014zwa}, and strengthening the analogy between VdW fluids and charged AdS black holes \cite{Kubiznak:2012wp,Wei:2015iwa,Wei:2019uqg,Wei:2019yvs}. A number of novel phase behaviors are also discovered in this subject, such as the reentrant
phase transition \cite{Gunasekaran:2012dq,Altamirano:2013ane}, triple points \cite{Altamirano:2013uqa, Wei:2014hba}, superfluid black holes \cite{Hennigar:2016xwd}, etc.

Nevertheless, the holographic interpretation of this extended thermodynamics is somewhat difficult to figure out \cite{Kastor:2014dra,Karch:2015rpa,Johnson:2014yja,Dolan:2014cja,Zhang:2014uoa,Zhang:2015ova,Dolan:2016jjc,McCarthy:2017amh}. Via the AdS/CFT correspondence, the thermodynamics in the bulk admits a description
of the dual CFT at finite
temperature. In particular, the mass $M$, entropy
$S$ and Hawking temperature $T$ for black holes can be directly mapped to the energy $E$, entropy $S$ and temperature $T$ of thermal states in the dual CFT \cite{Hawking:1982dh,Witten:1998zw}. However, there is no dual correlation between the bulk pressure $P$ and the pressure $p$ of the dual field theory, and the thermodynamic volume $V$ of black holes does not relate to the CFT volume $\mathcal{V} \sim l^{D-2}$ (supposing the curvature radius $R$ coincides with the AdS radius $l$) \cite{Johnson:2014yja}. On the other hand, the Smarr relation relating bulk quantities should correspond to the Euler relation for CFT quantities. While the latter relation contains no dimension dependent factors, unlike the former one.

These issues can be addressed by extending the parameter space of bulk thermodynamics. It has been argued that varying the bulk pressure $P$, is equivalent to varying the central charge $C$ or the number of colors $N$ in the dual gauge theory \cite{Kastor:2009wy,Kastor:2014dra,Karch:2015rpa,Johnson:2014yja,Dolan:2014cja}, via a holographic dictionary
\begin{equation}
C=k \frac{l^{D-2}}{G}, \label{eq:central}  
\end{equation}
where $k$ is a numerical factor (more details can be seen in Appendix \ref{central}). If the variation of Newton’s constant $G$ is included, the central charge $C$ can be fixed so that the field theory remains unchanged, and varying $\Lambda$ alters the CFT volume $\mathcal{V}$ more naturally. Inspired by this perspective, an exact match between the CFT thermodynamics and the extended gravitational thermodynamics has been made for charged (rotation) AdS black holes \cite{Visser:2021eqk}. The possible critical behaviors in the holographic dual of extended thermodynamics of the black holes were also studied \cite{Cong:2021jgb}. Three thermodynamic ensembles were found to exhibit intriguing critical behaviors, including the VdW like phase transition in fixed $(\tilde{Q},\mathcal{V},C)$ ensemble, the Hawking-Page like phase transition in fixed $(\tilde{\Phi},\mathcal{V},C)$ ensemble, as well as a new zeroth-order
phase transition between the high-entropy and low-entropy phases in fixed $(\tilde{Q},\mathcal{V}, \mu)$ ensemble, where $\tilde{Q}$, $\tilde{\Phi}$, $\mu$ denote the CFT electric charge, CFT electric potential and chemical potential for $C$ respectively.

A mixed bulk/boundary formalism of holographic thermodynamics was also proposed \cite{Cong:2021fnf}, where both the variations of bulk pressure $P$ and central charge $C$ are included in the first law of AdS black holes. Within this framework, we can study the bulk thermodynamics with the familiar pressure-volume term, meanwhile keeping the boundary central charge fixed. For charged AdS black holes, it was shown that a first-order phase transition between small and large black holes can occur when the dual field theory contains many degrees of freedom (a large central charge), whereas for the dual field theory with few degrees of freedom (a small central charge), there is no first-order phase transition in the bulk. Moreover, the associated critical central charge has a universal property---completely governed by the electric charge $Q$ and independent of the bulk pressure $P$.

Recently, this mixed bulk/boundary formalism was extended to the Born-Infeld AdS black holes \cite{Kumar:2022fyq}. Similar bulk small-large black hole phase transition was found when the boundary central charge exceeds a critical value. However, the universal property of the critical central charge breaks down, due to its dependence on the Born-Infeld parameter $b$ and the critical bulk pressure $P_c$. For a substantially
small $b$, there exists a no black hole region and a branch of large black hole solutions. These findings indicate that the non-linear electrodynamics has an important effect on the critical behavior of central charge. 

The aim of this paper is to search for another possible non-linear effect---reentrant phase transition in the holographic thermodynamics of Born-Infeld AdS black holes. This kind of large-small-large black hole phase transition, first discovered for $D=4$ charged Born-Infeld AdS black holes \cite{Gunasekaran:2012dq}, has gained great attention in the study on extended gravitational thermodynamics \cite{Altamirano:2013ane,Frassino:2014pha,Zou:2013owa,Altamirano:2014tva,Sherkatghanad:2014hda,Kubiznak:2015bya,Hennigar:2015wxa,Wei:2015ana,Hendi:2014kha,Hennigar:2015esa,KordZangeneh:2017lgs,Dehyadegari:2017flm,Dehyadegari:2017hvd,Dykaar:2017mba,Hennigar:2016ekz,Zou:2016sab,Xu:2019yub,Wang:2018xdz,Hendi:2017mfu,NaveenaKumara:2020biu,Dehghani:2020blz,Guo:2021ere,Ye:2022uuj,NingchenBai:2022ayc,Momennia:2021ktx,Astefanesei:2021vcp}. Within the new framework of holographic thermodynamics, the question arises whether such a phase transition can still be observed, both in the bulk and the dual CFT. In this work, we will provide an investigation on this.

The outline of our paper is as follows. In Sec. \ref{thermodynamics}, we give a brief introduction to the holographic thermodynamics of Born-Infeld AdS black holes. The thermodynamics in the bulk, CFT and mixed bulk/boundary formalism are discussed respectively. In Sec. \ref{topology}, the global (number) and local (stability) properties of critical points in the central charge criticality are investigated by use of topological analysis. The further study on phase transitions are presented in Sec. \ref{phase_tran}. Finally, we summarize and discuss our findings in Sec. \ref{conclu}.

\section{Holographic thermodynamics}\label{thermodynamics}

\subsection{Thermodynamics in the bulk}
The action of the Einstein-Born-Infeld theory in $D$-dimensional spacetime with a negative cosmological constant $\Lambda=-\frac{(D-1)(D-2)}{2 l^{2}}$ is given by \cite{Dey:2004yt,Cai:2004eh}
\begin{equation}
\mathcal{I}=\frac{1}{16 \pi G} \int d^{D} x \sqrt{-g}\bigl[ R-2 \Lambda +L(F)\bigr],
\label{eq:BIaction}
\end{equation}
and $L(F)$ reads
\begin{equation}
L (F)=4 b^{2}\left(1-\sqrt{1+\frac{F^{\mu \nu} F_{\mu \nu}}{2 b^{2}}}\right),
\end{equation}
in which the Born-Infeld
parameter $b$ represents the maximal electric field strength as the magnetic field vanishes. The electrodynamic field
tensor is $F_{\mu \nu}=\partial_{\mu} A_{\nu} - \partial_{\nu} A_{\mu}$. Here we use the usual conventions in the study of black hole thermodynamics, setting the coupling constant of the matter part  to be $1/16 \pi G$.

Consider  the black holes with static spherically symmetric  metric
\begin{equation}
d s^{2}=-f(r) d t^{2}+\frac{d r^{2}}{f(r)}+r^{2} d \Omega_{D-2}^{2},
\end{equation}
where $d \Omega_{D-2}^{2}$ is the line element of $(D-2)$-sphere with unit radius, and
\begin{equation}
\begin{aligned}
f(r)=& 1+\frac{r^{2}}{l^{2}}-\frac{m}{r^{D-3}} +\frac{4 b^{2} r^{2}}{(D-1)(D-2)}\left(1-\sqrt{1+\frac{(D-2)(D-3) q^{2}}{2 b^{2} r^{2 D-4}}}\right) \\
&+\frac{2 (D-2) q^{2}}{(D-1) r^{2 D-6}} \times{ }_{2} F_{1}\left[\frac{D-3}{2 D-4}, \frac{1}{2}, \frac{3 D-7}{2 D-4},-\frac{(D-2)(D-3) q^{2}}{2 b^{2} r^{2 D-4}}\right],
\end{aligned}
\end{equation}
with  hypergeometry function ${ }_{2} F_{1}$. The parameters $m$ and $q$ are related to total mass and charge of the black hole by \cite{Dey:2004yt,Cai:2004eh,Miskovic:2008ck}
\begin{equation}
M=\frac{(D-2) m \omega}{16 \pi G}, \quad Q=\frac{q \omega}{4 \pi G} \sqrt{\frac{(D-2)(D-3)}{2}}, \label{eq:charge}
\end{equation}
where $\omega=2 \pi^{\frac{D-1}{2}} / \Gamma[(D-1) / 2]$ is the volume of the unit $(D-2)$-sphere. Solving $f(r_+) = 0$, the black hole mass $M$ can be expressed as
\begin{equation}
\begin{aligned}
M=& \frac{(D-2) \omega r_{+}^{D-3}}{16 \pi G}\Biggl[1+\frac{r_{+}^{2}}{l^{2}}+\frac{4 b^{2} r_{+}^{2}}{(D-1)(D-2)}\left(1-\sqrt{1+\frac{16 \pi^{2} G^{2} Q^{2}}{b^{2} \omega^{2} r_{+}^{2 D-4}}}\right)\\
&+\frac{64 \pi^{2} G^{2} Q^{2}}{(D-1)(D-3) \omega^{2} r_{+}^{2 D-6}} \times{ }_{2} F_{1}\left[\frac{D-3}{2 D-4}, \frac{1}{2}, \frac{3 D-7}{2 D-4},-\frac{16 \pi^{2} G^{2} Q^{2}}{b^{2} \omega^{2} r_{+}^{2 D-4}}\right] \Biggr],
\end{aligned}
\label{eq:Mexp}
\end{equation}
where $r_+$ is the radius of the event horizon.
The gauge potential $A_{\nu}=(A_{t},0,0,\cdots)$ is identified by
\begin{equation}
\begin{aligned}
A_{t}=-\frac{4 \pi G Q}{(D-3) \omega r^{D-3}} \times{ }_{2} F_{1}\left[\frac{D-3}{2 D-4}, \frac{1}{2}, \frac{3 D-7}{2 D-4},-\frac{16 \pi^{2} G^2 Q^{2}}{b^{2} \omega^{2} r^{2 D-4}}\right].
\end{aligned}
\end{equation}
The associated electromagnetic potential, defined by $\Phi=\left.A_{\mu} \chi^{\mu}\right|_{r \rightarrow \infty}-\left.A_{\mu} \chi^{\mu}\right|_{r=r_{+}}$, where $\chi=\partial_{t}$ is the Killing vector, can be calculated as
\begin{equation}
\begin{aligned}
\Phi=\frac{4 \pi G Q}{(D-3) \omega r_{+}^{D-3}} \times{ }_{2} F_{1}\left[\frac{D-3}{2 D-4}, \frac{1}{2}, \frac{3 D-7}{2 D-4},-\frac{16 \pi^{2} G^2 Q^{2}}{b^{2} \omega^{2} r_{+}^{2 D-4}}\right]. \label{eq:phi}
\end{aligned}
\end{equation}
The entropy and Hawking temperature of the black hole are given by
\begin{align}
S &= \frac{A}{4 G}=\frac{\omega r_{+}^{D-2}}{4 G}, \\
T &= \frac{\kappa}{2 \pi}=\frac{1}{4 \pi}\Biggl[\frac{(D-1) r_{+}}{l^{2}}+\frac{D-3}{r_{+}} +\frac{4 b^{2} r_{+}}{D-2} \times \left(1-\sqrt{1+\frac{16 \pi^{2} G^2 Q^{2}}{b^{2} \omega^{2} r_{+}^{2 D-4}}}\right)\Biggr], \label{eq:tem}
\end{align}
where $A$ is the horizon area, and $\kappa$ is the surface gravity.

Identifying $A$, $Q$, $\Lambda$ and $b$ as independent thermodynamic quantities, by setting $D$-dimensional Newton's constant $G=1$, the  first law of thermodynamics and Smarr relation for Born-Infeld AdS black holes can be expressed as \cite{Yi-Huan:2010jnv,Gunasekaran:2012dq}
\begin{gather}
\delta M =\frac{\kappa}{8 \pi G} \delta A+\Phi \delta Q+\frac{\Theta}{8 \pi G} \delta \Lambda+\mathcal{B} \delta b, \label{eq:first_gravitational} \\
M =\frac{D-2}{D-3} \frac{\kappa A}{8 \pi G}+\Phi Q-\frac{1}{D-3}\left(\frac{\Theta \Lambda}{4 \pi G}+\mathcal{B} b\right), \label{eq:smarr_gravitational}
\end{gather}
where $\mathcal{B}$ is the thermodynamic quantities conjugate to $b$, which can be interpreted as the Born-Infeld vacuum polarization \cite{Gunasekaran:2012dq}. The $\Theta$ denotes the quantity conjugate to $\Lambda$, and can be defined in a geometric way in terms of surface integrals of the Killing potential \cite{Lemos:2018cfd}. 

If we further interpret the cosmological constant $\Lambda$ as the thermodynamic pressure by \cite{Caldarelli:1999xj,Kastor:2009wy,Dolan:2010ha}
\begin{eqnarray}
P=-\frac{\Lambda}{8 \pi G},
\end{eqnarray}
and assume that the Newton's constant $G$ is held fixed, the first law and Smarr relation can be cast into
\begin{gather}
\delta M = T \delta S +\Phi \delta Q+ V \delta P + \mathcal{B} \delta b, \\
M = \frac{D-2}{D-3} T S+\Phi Q-\frac{2}{D-3} V P-\frac{1}{D-3} \mathcal{B} b,
\end{gather}
where $V=-\Theta=\omega r^{D-1}_{+}/\left(D-1\right)$ can be identified as the thermodynamic volume of the black hole. In this context, the black hole mass $M$ can be identified as the thermodynamic enthalpy $H$ of the gravitational system, for the presence of a $V \delta P$ term. Studying the black hole thermodynamics including this term has been dubbed \textit{black hole chemistry} \cite{Kubiznak:2016qmn}, in which plenty of interesting phase transitions and phase behaviors are found. 

\subsection{Thermodynamics in the CFT}

As mentioned in the Introduction, the above thermodynamics in the bulk admits a description
of the dual CFT at finite
temperature, which can be achieved by including the variation of Newton's constant $G$. In this new parameter space, the Smarr relation Eq. (\ref{eq:smarr_gravitational}) holds true, but the first law Eq. (\ref{eq:first_gravitational}) needs to be corrected as follows, 

%We can assume the corrected first law takes the form
%\begin{eqnarray}
%\delta M = \frac{\kappa}{8 \pi G} \delta A +\Phi \delta Q+\frac{\Theta}{8 \pi G} \delta \Lambda +\mathcal{B} \delta b + X \delta G. \label{eq:first_corrected}
%\end{eqnarray}
%In this context, when considering a constant $G$, i.e. $\delta G=0$, the above relation reduces to Eq. (\ref{eq:first_gravitational}). To compute the the coefficient $X$, we can treat the mass $M$ as a function of $A$, $Q$, $\Lambda$, $b$ and $G$, i.e. $M=M(A,Q,\Lambda,b,G)$, and formally rewrite the first law as
%\begin{equation}
%\delta M = \frac{\partial M}{\partial A} \delta A + \frac{\partial M}{\partial Q} \delta Q + \frac{\partial M}{\partial \Lambda} \delta \Lambda+ \frac{\partial M}{\partial b} \delta b + \frac{\partial M}{\partial G} \delta G.
%\end{equation}
%These quantities scale as
%\begin{equation}
%[M]=[Q]=[b]=\frac{1}{L}, \quad [\Lambda]=\frac{1}{L^{2}}, \quad [G]=[A]=L^{D-2}.
%\end{equation}
%From Eq. (\ref{eq:first_corrected}) and scaling argument, we have
%\begin{equation}
%-M= (D-2) \frac{\kappa}{8 \pi G} A - \Phi Q -2 \frac{\Theta}{8 \pi G} \Lambda - \mathcal{B} b +(D-2) X G.
%\end{equation}
%Comparing this with the Smarr relation Eq. (\ref{eq:smarr_gravitational}), it is easy to read off $X=-(M- \Phi Q)/G$, and thus the corrected first law reads
\begin{equation}
\delta M = \frac{\kappa}{8 \pi G} \delta A + \Phi \delta Q + \frac{\Theta}{8 \pi G} \delta \Lambda + \mathcal{B} \delta b -\frac{(M - \Phi Q)}{G} \delta G, \label{first_corrected_final}
\end{equation}
which can be derived from Eq. 
(\ref{eq:Mexp}). The factor in front of  $\Phi Q$  differs from that presented previously \cite{Kumar:2022fyq} due to different definition of charge and potential. In Ref. \cite{Kumar:2022fyq}, the charge and potential are taken as $Q^{\prime}=\sqrt{G} Q$ and $\Phi^{\prime}=\Phi / \sqrt{G}$, which leads to $\Phi \delta Q + \Phi Q \delta G/G=\Phi^{\prime} \delta Q^{\prime} + \frac{1}{2} \Phi^{\prime} Q^{\prime} \delta G/G$.

Eq. (\ref{first_corrected_final}) can also be cast into
\begin{eqnarray}
\delta M = \frac{\kappa}{8 \pi} \delta \left(\frac{A}{G}\right) + \Phi \delta Q + \frac{\Theta}{8 \pi} \delta \left(\frac{\Lambda}{G}\right) + \mathcal{B} \delta b - \left(M -  \frac{\kappa A}{8 \pi G} - \Phi Q - \frac{\Theta \Lambda}{8 \pi G} \right) \frac{\delta G}{G}. \label{eq:first_half}
\end{eqnarray}
The first term can be identified with the $T \delta S$ term, and the third term can be treated as the $V \delta P$ term, while the last term does not directly correspond to a  thermodynamic interpretation. To obtain a first law with a well-defined thermodynamic correspondence, one can further write Eq. (\ref{eq:first_half}) as
\begin{eqnarray}
\delta M = \frac{\kappa}{2 \pi} \delta \left(\frac{A}{4 G}\right)+\frac{\Phi}{l} \delta(Q l)-\frac{M}{D-2} \frac{\delta l^{D-2}}{l^{D-2}} + \frac{\mathcal{B}}{l} \delta (b l)+\left(M-\frac{\kappa A}{8 \pi G}-\Phi Q\right) \frac{\delta\left(l^{D-2} / G\right)}{l^{D-2} / G}, \label{eq:first_half_half}
\end{eqnarray}
where $\delta \Lambda/ \Lambda = - 2 \delta l/l$ has been used. After converting the bulk field strength in action (\ref{eq:BIaction}) to a canonical normalized field strength
of dimension $2$ (mass unit) \cite{Chamblin:1999tk,Karch:2015rpa}, the holographic dictionary can be identified as
\begin{gather}
E=M, \quad \tilde{\Phi}=\Phi / l, \quad \tilde{Q}=Q l, \quad \tilde{\mathcal{B}}= \mathcal{B}/l, \quad \tilde{b}=b l, \quad \mathcal{V} \sim l^{D-2}, \quad C \sim l^{D-2} / G. \label{eq:dic}
\end{gather}
Inserting them into Eq. (\ref{eq:first_half_half}) gives
\begin{eqnarray}
\delta E=T \delta S+\tilde{\Phi} \delta \tilde{Q} -p \delta \mathcal{V} + \tilde{\mathcal{B}} \delta \tilde{b} + \mu \delta C, \label{eq:first_CFT}
\end{eqnarray}
with
\begin{eqnarray}
p=\frac{E}{(d-2) \mathcal{V}}, \quad \mu = \frac{1}{C} \left(E-T S - \tilde{\Phi} \tilde{Q} \right),
\end{eqnarray}
in which every term has a direct thermodynamic interpretation. The quantity $p$ can be identified as the field theory pressure that satisfies the CFT equation of state
\begin{eqnarray}
E=(D-2) p \mathcal{V},
\end{eqnarray}
and $\mu$ can be identified as the chemical potential  for the central charge $C$. Since all the quantities are defined on the boundary, Eq. (\ref{eq:first_CFT}) can be treated as the thermodynamic first law in the dual CFT of Born-Infeld AdS black holes. The holographic Euler equation can also be given by rearranging the expression for $\mu$:
\begin{eqnarray}
E=TS + \tilde{\Phi} \tilde{Q} + \mu C, \label{eq:Euler}
\end{eqnarray}
which is consistent with on-shell calculation of grand canonical free energy $W = E - T S - \tilde{\Phi} \tilde{Q} \equiv \mu C$ in the holographic field theory \cite{Miskovic:2008ck,Fernando:2006gh}. In contrast to the Smarr relation Eq. (\ref{eq:smarr_gravitational}), the Euler equation for CFT does not contain any dimension dependent factors, just as expected.

Moreover, it is easy to verfy that the CFT first law Eq. (\ref{eq:first_CFT}) and Euler equation Eq. (\ref{eq:Euler}) hold ture even when the boundary curvature radius $R$ is unequal to the AdS radius $l$, with the holographic dictionary redefined as \cite{Savonije:2001nd,Karch:2015rpa,Visser:2021eqk,Cong:2021jgb}
\begin{gather}
E=M \frac{l}{R}, \quad T=\frac{\kappa}{2 \pi} \frac{l}{R}, \quad S=\frac{A}{4 G}, \nonumber \\
\tilde{\Phi}=\frac{\Phi}{l} \frac{l}{R}, \quad \tilde{Q}=Q l, \quad \tilde{\mathcal{B}}= \frac{\mathcal{B}}{l} \frac{l}{R}, \quad \tilde{b}=b l, \quad \mathcal{V} \sim R^{D-2}, \quad C \sim l^{D-2} / G. \label{eq:dic_redefined}
\end{gather}
In this context, the CFT volume $\mathcal{V}$ and central charge $C$ are now completely independent.

\subsection{Mixed bulk/boundary formalism}
As presented in Ref. \cite{Cong:2021fnf}, a “mixed” bulk/boundary form of first law can also be constructed, which admits both the variation of bulk pressure $P$ and boundary central charge $C$. This is accomplished by inserting
\begin{equation}
\frac{\delta G}{G}=-\frac{2}{D} \frac{\delta C}{C}-\frac{D-2}{D} \frac{\delta P}{P},
\label{eq:delta_G}
\end{equation}
following from Eq. (\ref{eq:central}), into Eq. (\ref{eq:first_half}). The result shows that
\begin{equation}
\delta M=T \delta S+\Phi \delta Q + \mathcal{B} \delta b +V_{C} \delta P+\mu_C \delta C, \label{eq:first_mixed}
\end{equation}
where 
\begin{equation}
V_C=\frac{M - \Phi Q - \mathcal{B} b}{D P}, \quad \mu_C=\frac{2 P(V_C-V)}{C(D-2)}
\end{equation}
are the new thermodynamic volume and chemical
potential, respectively. This mixed form of first law enables us to investigate the bulk thermodynamics with the familiar pressure-volume term, meanwhile keeping the boundary central charge fixed. Notice that the difference in $V_C$ with that obtained from Ref. \cite{Kumar:2022fyq} is due to the different definition of electric charge and potential.

In the followings, we first investigate the critical behaviors in the mixed bulk/boundary formalism, and specially, in the fixed $(Q,P,b,C)$ ensemble. The possible critical behaviors in the dual CFT will be examined in the last section. The associated temperature from Eq. (\ref{eq:tem}) can be written as
\begin{align}
T(r_+,C,z^i)=&\frac{1}{4 \pi } \Bigl[\frac{D-3}{r_+} -\frac{4 b^2 r_+ \left(x-1\right)}{D-2}+ (16 \pi)^{2/D} (D-1) r_+ C^{-2/D} y^{-2/D}\Bigr], \label{eq:tem_mixed}
\end{align}
where $z^i=(Q,P,b)$ are the bulk parameters and
\begin{align}
& x=\sqrt{1+\frac{16^{\frac{4}{D}-1} \pi ^{4/D} y^{2-\frac{4}{D}} r_+^{4-2 D} C^{-4/D} Q^2}{b^2 \omega ^2}},\\
& y=\frac{(D-2) (D-1)}{P}, \quad \omega=\frac{2 \pi ^{\frac{D-1}{2}}}{\Gamma \left(\frac{D-1}{2}\right)}.
\end{align}
The definition of bulk pressure $P=-\Lambda/8\pi G$ and CFT central charge $C=k l^{D-2}/G$ has been used to transform $(l, G)$ to $(P, C)$, and for simplicity we take $k=1$. 

The corresponding thermodynamic critical points are given by
\begin{equation}
\left(\frac{\partial T}{\partial r_+}\right)_{C, z^i}=0, \quad\left(\frac{\partial^{2} T}{\partial r_+^{2}}\right)_{C, z^i}=0, \label{eq:con}
\end{equation}
and more concretely,
\begin{align}
&\left(\frac{\partial T}{\partial r_+}\right)_{C, z^i}=\frac{1}{4 \pi } \Bigl[\frac{3-D}{r_+^2}-\frac{4 b^2 (x-1)}{D-2} +\frac{2^{\frac{16}{D}-2} \pi ^{\frac{4}{D}} y^{2-\frac{4}{D}} r_+^{4-2 D} C^{-\frac{4}{D}} Q^2 }{x \omega ^2}+(D-1) (16 \pi )^{\frac{2}{D}} y^{-\frac{2}{D}} C^{-\frac{2}{D}}\Bigr], \label{eq:first_con} \\
&\left(\frac{\partial^{2} T}{\partial r_+^{2}}\right)_{C, z^i}=\frac{D-3}{2 \pi  r_+^3}-\frac{2^{\frac{16}{D}+4} (2 D-5) \pi ^{4/D} \omega ^2 b^2 y^2 r_+^3 Q^2 C^{4/D}+2^{32/D} (D-3) \pi ^{8/D} y^{4-\frac{4}{D}} r_+^{7-2 D} Q^4}{\pi  x \omega ^2 \left(2^{\frac{16}{D}+4} \pi ^{4/D} y^2 r_+^4 Q^2 C^{4/D} + 2^8 b^2 \omega ^2 y^{4/D} r_+^{2 d} C^{8/D}\right)}. \label{eq:second_con}
\end{align}
The free energy $F(T,C,z^i)$ in this ensemble can be identified as $F=M-T S$, i.e.,
\begin{align}
&F=\frac{\omega  r_+^{D-3} P}{(D-1)^2 (D-2)^2} \Bigl[(16 \pi )^{-2/D} C^{2/D} y^{2/D}
\bigl(4 b^2 r_+^2 (x-1)+D^2-3 D+2\bigr)-\left(D^2-3 D+2\right) r_+^2\Bigr] \nonumber \\
& \quad \quad +\frac{2^{\frac{8}{D}-2} (D-2)^2 \pi ^{2/D}}{(D-3) P \omega} y^{-2/D} r_+^{3-D} C^{-2/D} Q^2 \times{ }_{2} F_{1}\Bigl[\frac{D-3}{2 D-4},\frac{1}{2},\frac{3 D-7}{2 D-4},-\frac{16^{\frac{4}{D}-1} \pi ^{4/D} y^{2-\frac{4}{D}} Q^2}{b^2 \omega ^2 r_+^{2 D - 4} C^{4/D} }\Bigr], \label{eq:free}
\end{align}
where $r_+=r_+(T,C,z^i)$ can be obtained from Eq. (\ref{eq:tem_mixed}). The stable phase corresponds to the global minimum of $F(T,C,z^i)$ for its fixed parameters $T$, $C$ and $z^i$. This further allows us to investigate phase transition behaviors between different phases.

\section{Topological analysis for critical points}\label{topology}
Due to the complexity of the Hawking temperature Eq. (\ref{eq:tem_mixed}), it is difficult to obtain an analytical solution for critical point condition Eq. (\ref{eq:con}). In $D=4$ case, an approximating solution is achieved by expanding the condition upto $\mathcal{O} (1/b^{2})$ \cite{Kumar:2022fyq}. For our purpose, we aim to search for the possible reentrant phase transition in the mixed bulk/boundary formalism, and thus the information for critical points in different range of parameters is required.

Topological analysis can be used to address this challenge. Some useful information can be directly obtained without exact calculation of critical points, such as the existence and number (odd or even) of critical points, as well as the possible transition in phase structures, which would be quite helpful for the investigation on black hole thermodynamics \cite{Wei:2021vdx,Yerra:2022alz,Yerra:2022coh,Yerra:2022eov,Bai:2022klw}. Such an approach relies on a topological quantity \textit{topological charge} \cite{Duan:2018rbd,Duan:1984ws,Fu:2000pb,Cunha:2017qtt}, or equally, the \textit{Brouwer degree} \cite{dinca2021brouwer}. In the followings, we will give a brief introduction to this.

Consider an open and bounded set $X \subset \mathbb{R}^n$ with a (at least) $C^1$-smooth map $f: X \rightarrow \mathbb{R}^n$. Let $y \in f \backslash f(\partial X)$ be a regular value of $f$, then the set $f^{-1} (y)=\left\{x_1,x_2,\cdots\right\}$ with $x_n \in X$ has a finite number of points, such that $f(x_n)=y$. Suppose the Jacobian $J(x_n)=\text{det} (\partial f/ \partial x_n) \neq 0$, one can define a topological quantity, called the Brouwer
degree of the map \cite{dinca2021brouwer}
\begin{equation}
\operatorname{deg}(f,X,y)=\sum_{x_n \in f^{-1}(y)} \operatorname{sgn} J(x_n), \label{eq:Brouwer_degree}
\end{equation}
where sgn denotes the sign function. This quantity is a topological characteristic of the map itself, which does not depend on the choice of the regular value $y$ and remains constant under continuous deformations of the map. 

%%%%%%%%%%%%%%%%%%%%
\begin{figure}
\centering
\includegraphics[height=5cm]{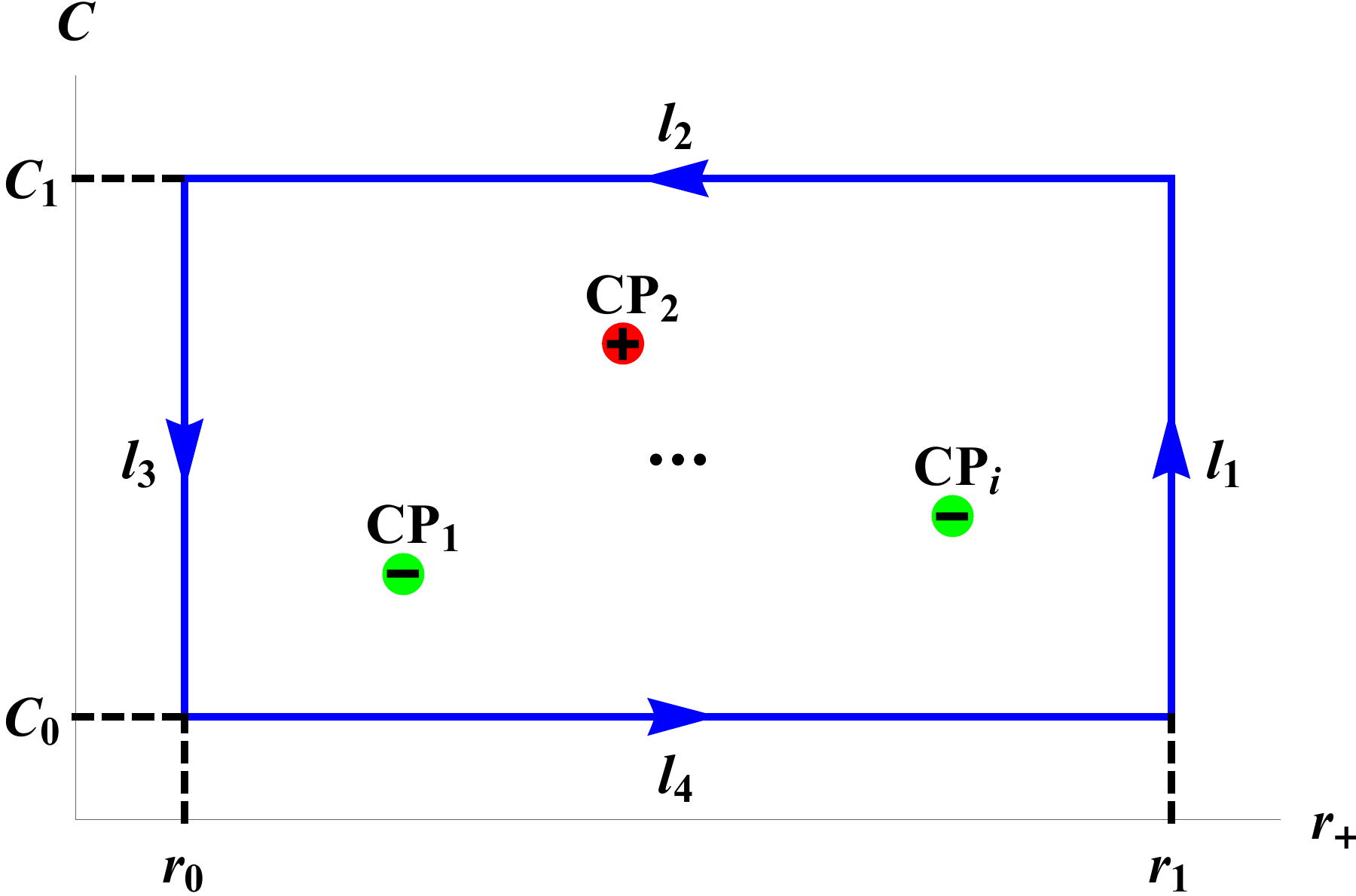}
\caption{\label{fig:contour}Sketch map of the contour $L=\sum_{i} \cup l_{i}$ on the $(r_+, C)$ plane. Arrows refer to the positive orientation.}
\end{figure}
%%%%%%%%%%%%%%%%%%%%%%%%

To apply this tool to the study of thermodynamics, the function $f$ and its zero points need to be endowed with specific physical significance. In general case, one can use the first equation of condition Eq. (\ref{eq:con}) to eliminate one parameter in $T$, such as the central charge $C$ in our case, and then construct a two-dimensional vector field \cite{Wei:2021vdx} or one-dimensional function \cite{Bai:2022klw} utilizing this new $T$ to study the topological properties of critical points. However, due to the complexity of $(\partial_{r_+} T)_{z^i}$ in Eq. (\ref{eq:first_con}), it is difficult to achieve this. Note that $C$ also appears in $x$.

In order to overcome this problem, we can directly let $(\partial_{r_+} T)_{z^i} \equiv \Phi_{1}$ and $(\partial^2_{r_+} T)_{z^i} \equiv \Phi_{2}$ be the components of a two-dimensional vector field
\begin{equation}
\Phi=\left(\Phi_{1}, \Phi_{2}\right),
\end{equation}
and then zero points of $\Phi$ automatically become critical points of the thermodynamic system. In this context, we can endow a topological charge for each critical point, and a total topological charge for the system to investigate the global topological properties of the Born-Infeld AdS black hole.

\subsection{Global properties}

For a two-dimensional vector field $\left(\Phi_{1}, \Phi_{2}\right)$, it can be shown that the topological charge (degree) for each zero point is equal to the winding number \cite{milnor1997topology}
\begin{equation}
w_{i} \equiv \frac{1}{2 \pi} \oint_{C_{i}} d \Omega,
\end{equation}
where $\Omega=\arctan \left(\Phi_{2} / \Phi_{1}\right)$ is the deflection angle of vector field, and $C_i$ is a smooth and positive oriented closed curve which encloses the $i$-th zero point. 

To calculate the total topological charge, we construct a contour enclosing all possible zero points. The total topological charge can be directly obtained by counting the variation of the deflection angle $\Omega$ of the vector field along the contour. Considering the central charge criticality, the contour we used is constructed on the $(r_+,C)$ plane, as shown in Fig. \ref{fig:contour}: $L=\sum_{i} \cup l_{i}=\{r_+=r_1, C_0 \leq C \leq C_1\}\cup\{C=C_1, r_0\leq r_+ \leq r_1\}\cup\{r_+=r_0, C_0\leq C \leq C_1\}\cup\{C=C_0, r_0\leq r_+ \leq r_1\}$.
Without loss of generality, we set $r_0=C_0=\delta$ and $r_1=C_1=\delta^{-1}$, then the total topological charge can be obtained as
\begin{equation}
Q= \frac{1}{2 \pi} \lim _{\delta \rightarrow 0^{+}} \sum_{i} \Delta \Omega_{l_i}. \label{eq:totalcharge}
\end{equation}
In order to conveniently discuss the behavior of vector field on the contour $L$, we also define
\begin{equation}
(r_+)_{\alpha} \equiv \delta^{-\alpha}, \quad C_{\beta} \equiv \delta^{-\beta},
\end{equation}
where $\alpha,\beta \in [-1,1]$. In this context, we can map the region enclosed by the contour to a finite region $(\alpha, \beta)$, and a smaller $\alpha$ or $\beta$ corresponds to a smaller $r_+$ or $C$.

Now we first examine the behavior of vector field $\Phi$ on the contour in $D=4$ case. The Hawking temperature $T$ in Eq. (\ref{eq:tem_mixed}) reduces to
\begin{equation}
T=\frac{1}{4\pi} \left[\frac{1}{r_+} + 2 r_+ \sqrt{\frac{6 \pi  P}{C}} -2 b^2 r_+ \left(x-1\right)\right],
\end{equation}
where
\begin{equation}
x=\sqrt{1+\frac{3 Q^2}{8 \pi  b^2 r_+^4 P C}},
\end{equation}
and the components of vector field reads
\begin{align}
&\Phi_1 = \frac{1}{4 \pi }\left[2 \sqrt{\frac{6 \pi  P}{C}} -\frac{1}{r_+^2} -2 b^2 (x-1) + \frac{3 Q^2}{2 \pi x r_+^4 P C}\right], \\
&\Phi_2 = \frac{1}{2 \pi  r_+^3}-\frac{9 Q^2 \left(8 \pi  b^2 r_+^4 P C +Q^2\right)}{8 \pi ^2 x r_+^5 P C \left(8 \pi  b^2 r_+^4 P C +3 Q^2\right)}.
\end{align}
Along $l_{1}$, we have $r_+=\delta^{-1}$, $C=\delta^{-\beta}$, and
\begin{align}
x (\delta \rightarrow 0^+) \sim 1 + \frac{3 Q^2}{16 \pi  b^2 P} \delta^{\beta + 4}
\end{align}
for $\beta \in [-1,1]$, which results in
\begin{align}
\Phi_1 (\delta \rightarrow 0^+) \sim \sqrt{\frac{3 P}{2 \pi}} \delta^{\beta/2}, \quad \Phi_2 (\delta \rightarrow 0^+) \sim \frac{\delta ^3}{2 \pi }.
\end{align}
Since $|\Phi_2/\Phi_1| \rightarrow 0$ and $\Phi_1 > 0$, the vector 
$\Phi$ is horizontal to the right on $l_1$, i.e., $\Omega_{l_1}=0$. This indicates that
\begin{eqnarray}
 \Delta\Omega_{l_1}=0.
\end{eqnarray}
Along $l_{2}$, we have $r_+=\delta^{-\alpha}$, $C=\delta^{-1}$, and
\begin{align}
 x (\delta \rightarrow 0^+) \sim \left\{\begin{array}{ll}
\left[\frac{3 Q^2}{8 \pi  b^2 P}\right]^{1/2} \delta ^{2 \alpha +1/2} & \text { for } \alpha \in [-1, - 1/4), \\
\left[1+\frac{3 Q^2}{8 \pi b^2 P}\right]^{1/2} & \text { for } \alpha = - 1/4, \\
1 + \frac{3 Q^2}{16 \pi b^2  P} \delta ^{4 \alpha +1} & \text { for } \alpha \in (- 1/4, 1],
\end{array}\right.
\end{align}
which gives
\begin{align}
&\Phi_1 (\delta \rightarrow 0^+) \sim \left\{\begin{array}{ll}
-\frac{\delta ^{2 \alpha }}{4 \pi } & \text { for } \alpha \in [-1, 1/4), \\
-\frac{1-2 \sqrt{6 \pi  P}}{4 \pi } \delta^{1/2} & \text { for } \alpha = 1/4, \\
\sqrt{\frac{3 P}{2 \pi }} \delta^{1/2} & \text { for } \alpha \in (1/4, 1],
\end{array}\right. \\
& \Phi_2 (\delta \rightarrow 0^+) \sim \frac{\delta ^ {3 \alpha}}{2 \pi } \quad \text { for } \alpha \in [-1, 1].
\end{align}
Varying $r_+$ from $+ \infty$ to $0$ (or $\alpha$ from $1$ to $-1$), $\Phi_2$ is always positive whereas $\Phi_1$ varies from positive to negative, and $|\Phi_2/\Phi_1|$ changes from $0$ to $+ \infty$, suggesting that $\Omega$ changes $\pi/2$ in anti-clockwise direction along $l_2$, i.e.,
\begin{eqnarray}
\Delta\Omega_{l_2} &=& \frac{\pi}{2}.
\end{eqnarray}
Along $l_{3}$, we have $r_+=\delta$, $C=\delta^{-\beta}$, and
\begin{eqnarray}
x (\delta \rightarrow 0^+) &\sim& \left(\frac{3 Q^2}{8 \pi  b^2 P}\right)^{1/2} \delta ^{\beta/2 - 2}
\end{eqnarray}
for $\beta \in [-1, 1]$, which indicates that
\begin{align}
&\Phi_1 (\delta \rightarrow 0^+) \sim \left\{\begin{array}{ll}
\frac{3 b Q \delta ^{\beta /2}-(6 \pi  P)^{1/2}}{4 \pi (6 \pi  P)^{1/2} \delta ^2} & \text { for } \beta = [-1, 1] \backslash \beta_*, \\
\frac{b^2+(6 \pi  P)^{1/2} \delta ^{\beta_* /2}}{2 \pi } & \text { for } \beta = \beta_*,
\end{array}\right. \\
&\Phi_2 (\delta \rightarrow 0^+) \sim \left\{\begin{array}{ll}
\frac{(6 \pi  P)^{1/2}-3 b Q \delta ^{\beta /2}}{2 \pi (6 \pi  P)^{1/2} \delta ^3} & \text { for } \beta = [-1, 1] \backslash \beta_*, \\
-\frac{b^3 (6 \pi  P)^{1/2} \delta ^{1-\beta_*/2}}{\pi  Q} & \text { for } \beta = \beta_*,
\end{array}\right.
\end{align}
Here $\beta_*=-\log_{\delta} \left(3 b^2 Q^2/2 \pi P\right) \in (-1,1)$, corresponding to the central charge $C_*=3 b^2 Q^2/2 \pi P$, is the zero point of leading terms of $\Phi_1$ and $\Phi_2$. It is easy to see that from top to bottom, $\Phi_1$ changes from negative to positive at $C_*$, whereas $\Phi_2$ changes in the opposite direction. Moreover, for $\beta \in [-1, \beta_*) \cup (\beta_*, 1]$, $|\Phi_2/\Phi_1| \rightarrow + \infty$, thus the vector $\Phi$ points up and down vertically for $C > C_*$ and $C < C_*$, respectively. While for $\beta = \beta_*$, $|\Phi_2/\Phi_1| \rightarrow 0$, and $\Phi_1$ is positive, indicating that the vector $\Phi$ is horizontal to the right at $C_*$. These demonstrate that $\Omega$ changes $\pi$ in clockwise direction along $l_{3}$, i.e.,
\begin{eqnarray}
\Delta\Omega_{l_3}=- \pi.
\end{eqnarray}
Along $l_{4}$, we have $r_+=\delta^{-\alpha}$, $C=\delta$, and
\begin{align}
 x (\delta \rightarrow 0^+) \sim \left\{\begin{array}{ll}
\left[\frac{3 Q^2}{8 \pi  b^2 P}\right]^{1/2} \delta ^{2 \alpha -1/2} & \text { for } \alpha \in [-1, 1/4), \\
\left[1+\frac{3 Q^2}{8 \pi b^2 P}\right]^{1/2} & \text { for } \alpha = 1/4, \\
1 + \frac{3 Q^2}{16 \pi b^2  P} \delta ^{4 \alpha -1} & \text { for } \alpha \in (1/4, 1],
\end{array}\right.
\end{align}
which results in
\begin{align}
&\Phi_1 (\delta \rightarrow 0^+) \propto \left\{\begin{array}{ll}
\delta ^{2 \alpha -1/2} & \text { for } \alpha \in [-1, -1/8), \\
\delta^{-1/2} & \text { for } \alpha = [-1/8, 1],
\end{array}\right. \nonumber \\
\\
&\Phi_2 (\delta \rightarrow 0^+) \propto \left\{\begin{array}{ll}
- \delta ^{3 \alpha -1/2} & \text { for } \alpha \in [-1, 1/4), \\
- \delta^{5 \alpha - 1} & \text { for } \alpha \in (1/4, 1/2), \\
(4 \pi - 9 Q^2/P) \delta^{3/2} & \text { for } \alpha = 1/2, \\
\delta ^{3 \alpha } & \text { for } \alpha \in (1/2, 1],
\end{array}\right.
\end{align}
Varying $r_+$ from $0$ to $\infty$ (or $\alpha$ from $-1$ to $1$), $\Phi_1$ is always positive whereas $\Phi_2$ varies from negative to positive, and $|\Phi_2/\Phi_1|$ changes from $+ \infty$ to $0$, suggesting that $\Omega$ changes $\pi/2$ in anti-clockwise direction along $l_4$, i.e.,
\begin{eqnarray}
\Delta\Omega_{l_4}=\pi/2.
\end{eqnarray}
Then total topological charge for $D=4$ case can be calculated as
\begin{eqnarray}
Q_{D=4}=\frac{1}{2 \pi} \sum_{i} \Delta \Omega_{l_i}=0.
\end{eqnarray}
Note that this result does not depend on the values of bulk parameters $z^i=(Q,P,b)>0$. Since the well-defined topological charge for each critical point can only take $-1$ or $+1$, this result also suggests that in the central charge criticality, the $D=4$ charged Born-Infeld AdS black hole always possesses 
an even number of critical points (including $0$). This is very different from the $P-V$ criticality of the black hole \cite{Gunasekaran:2012dq}, in which even or odd number of critical points can exist in different ranges of parameters. Moreover, one of features of the reentrant phase structure is that it generally contains even (more concretely, two) critical points, and thus there is a high probability of finding a reentrant phase transition in the central charge criticality of the black hole.

A similar analysis provided in Appendix \ref{higher_d} shows that the total topological charge for arbitrary $D>4$ black hole is
\begin{eqnarray}
Q_{D>4}=+1.
\end{eqnarray}
We can justify that there is at least one critical point in the central charge criticality of $D>4$ black hole (otherwise a zero topological charge will be
obtained), and an odd number of critical points always exist, independent of the bulk parameters $z^i=(Q,P,b)>0$. Considering that the VdW like phase structure and the phase structure with triple point typically contain an odd number of critical points, there is a high probability of finding them in the $D>4$ black hole.

%%%%%%%%%%%%%%%%%%%%%%%%%%%
\begin{figure}
\center{\subfigure{\label{vecotr_4}
\includegraphics[height=6.5cm]{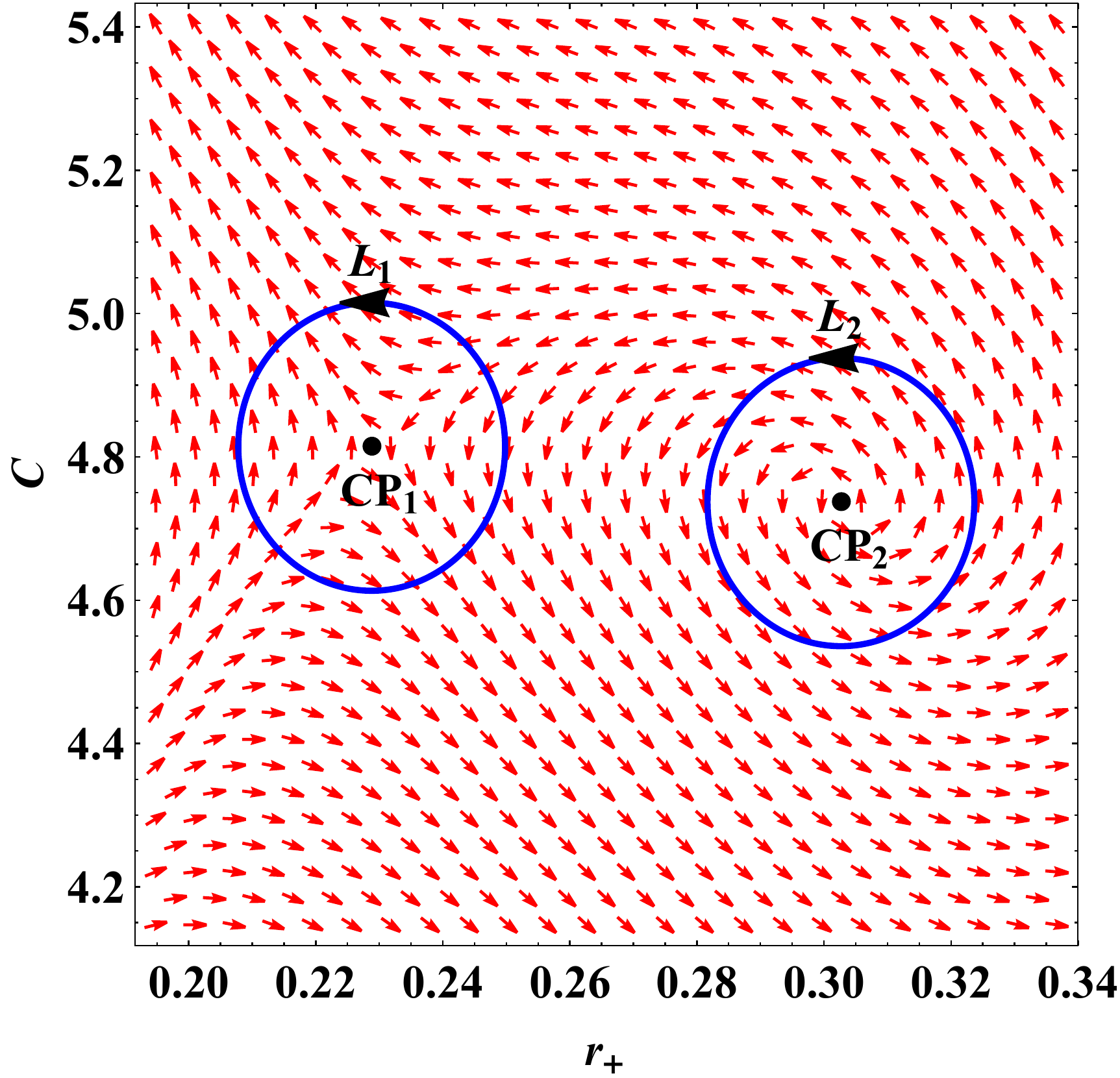}}
\hspace{1cm}
\subfigure{\label{vector_5}
\includegraphics[height=6.5cm]{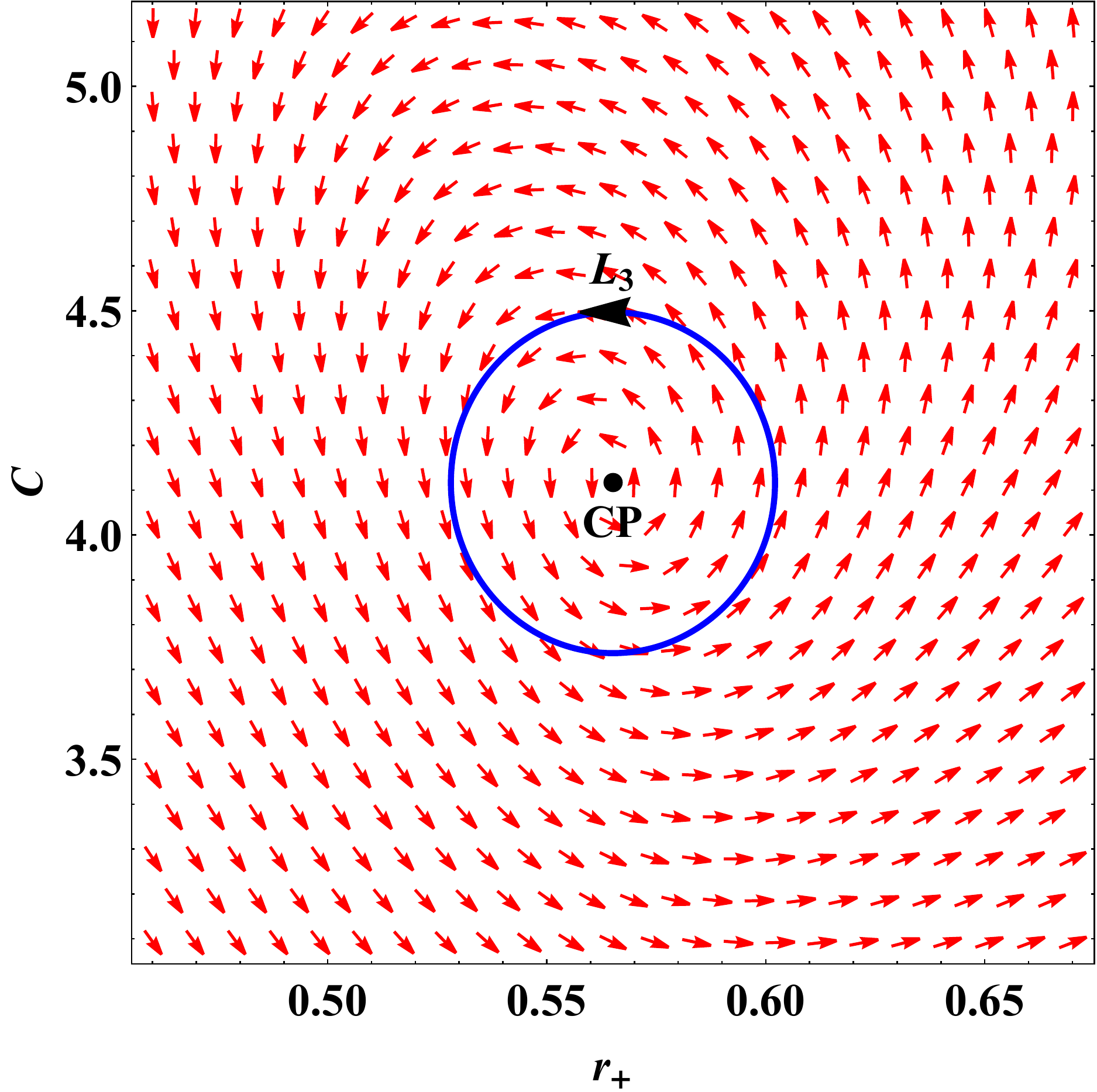}}}
\center{\subfigure{\label{omega_4}
\includegraphics[width=6.85cm]{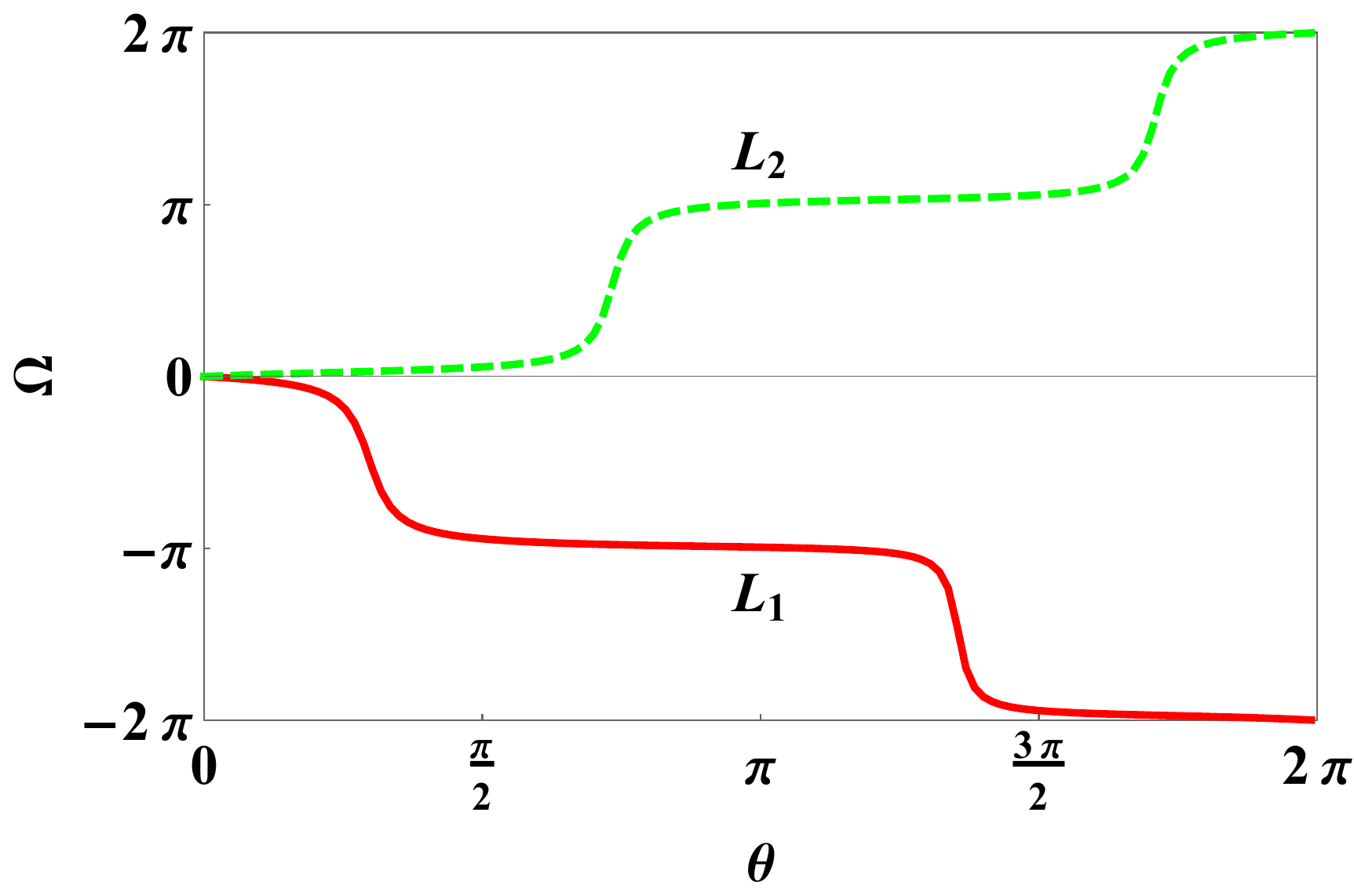}}
\hspace{1.1cm}
\subfigure{\label{omega_5}
\includegraphics[width=6.65cm]{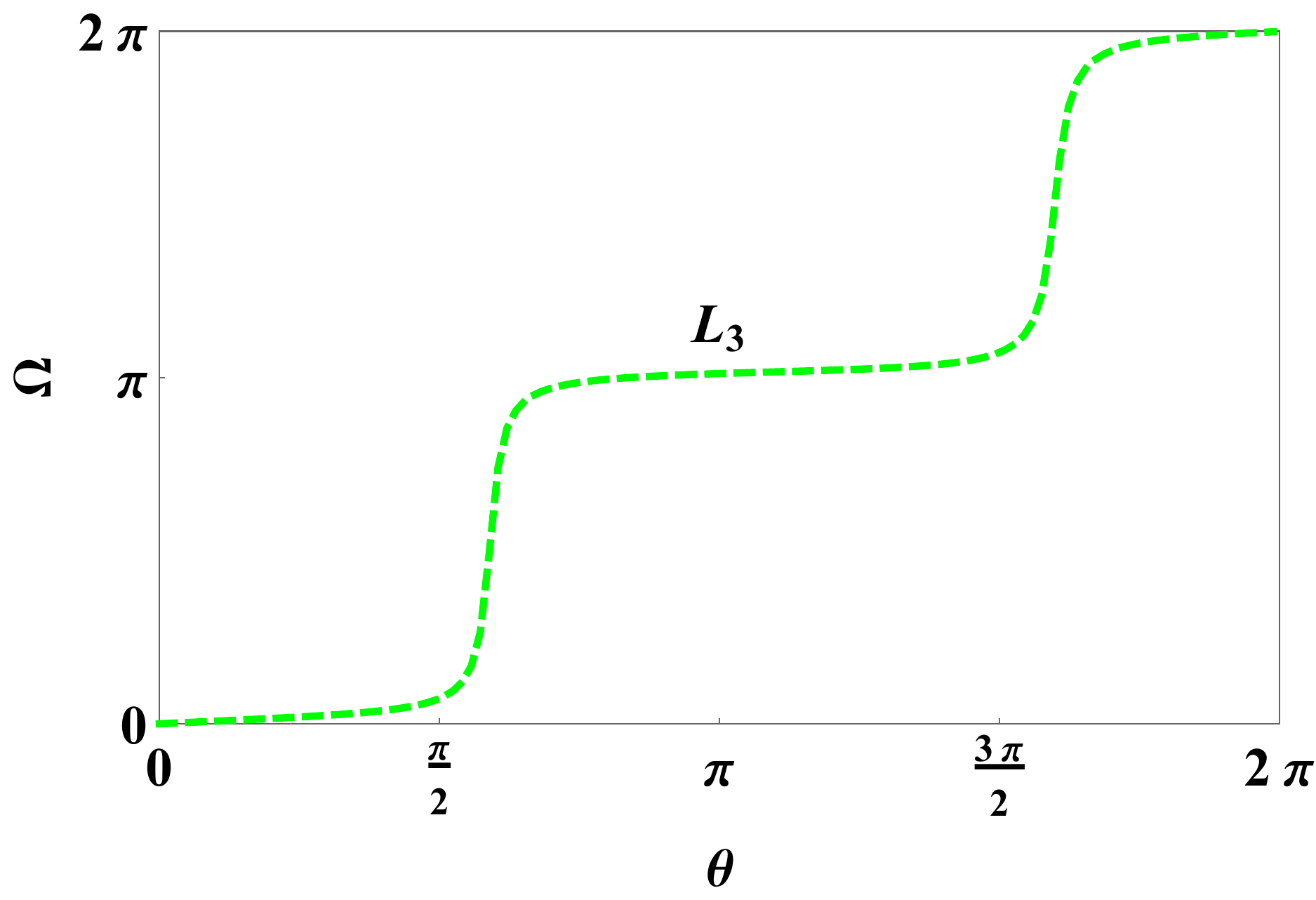}}}
\caption{\label{d=4} The normalized vector field in the $r_+-C$ diagrams (top) and $\Omega$ vs $\theta$ curves for different contours (bottom). Left: $D=4$ case. Right: $D=5$ case. We have set $(Q,P,b)=(1,1,2.3)$.}
\end{figure}
%%%%%%%%%%%%%%%%%%%%%%%%%%%

Taking $(Q,P,b)=(1,1,2.3)$ for example, we numerically plot the vector field in Fig. \ref{d=4}. For the $D=4$ case, there are two zero points $\text{CP}_1$ and $\text{CP}_2$ in the vector field, possessing opposite topological charges $-1$ and $+1$, and thus the total topological charge is $Q=Q_{\text{CP}_1} + Q_{\text{CP}_2}=0$, just as expected. For the $D=5$ case, only one critical point with topological charge $Q_{\text{CP}}=+1$ exists in the vector field, and the total topological charge indeed takes $Q=Q_{\text{CP}}=+1$. 

\subsection{Local properties}

In the above section, we have showed the global (topological) properties of critical points for charged Born-Infeld black holes in different dimensions. Now we display the local properties of the critical points with different topological charge $-1$ or $+1$. The topological charge is determined by the sign of Jacobian at zero points $x_i$,
\begin{eqnarray}
J(x_i)=\left|\begin{array}{ll}
(\partial_{r_+} \Phi_{1})_{x_i} & (\partial_{C} \Phi_{1})_{x_i} \\
(\partial_{r_+} \Phi_{2})_{x_i} & (\partial_{C} \Phi_{2})_{x_i}
\end{array}\right|.
\end{eqnarray}
At zero points, we have $(\partial_{r_+} \Phi_{1})_{x_i}=\Phi_{2}(x_i)=0$, thus the above determinant reduces to
\begin{eqnarray}
J(x_i)=-(\partial_{C} \Phi_{1})_{x_i} (\partial_{r_+} \Phi_{2})_{x_i}.
\end{eqnarray}
Using Eq. (\ref{eq:first_con}) and the definition of $\Phi_1$, we find
\begin{eqnarray}
\partial_{C} \Phi_{1}=-\frac{A}{B},
\end{eqnarray}
where
\begin{align}
A=&2^{\frac{8}{D}-1} \pi ^{\frac{2}{D}-1} r_+^{-2 D} y^{-4/D} C^{-\frac{D+4}{D}} \times \bigl[ 2^{\frac{8}{D}+4} (2 D-5) \pi ^{2/D} \omega ^2 b^2 r_+^{2 D+4} y^{\frac{4}{D}+2} C^{4/D} Q^2 \nonumber \\
&+2^{\frac{16}{D}+2} (D-1) (D-2) \pi ^{4/D} \omega ^2 x r_+^{2 D+4} y^{\frac{2}{D}+2} C^{2/D} Q^2 \nonumber \\
&+64 (D-1) (D-2) \omega ^4 b^2 x r_+^{4 D} y^{6/D} C^{6/D} +2^{24/D} (D-3) \pi ^{6/D} r_+^8 y^4 Q^4\bigr],\\
B=&4 D (D-2) x \bigl(2^{16/D} \pi ^{4/D} \omega ^2 y^2 r_+^4 Q^2 +16 b^2 \omega ^4 r_+^{2 D} y^{4/D} C^{4/D}\bigr).
\end{align}
For positive $(b, P, C, Q)$ and $D \geq 4$, $A, B>0$, $\partial_{C} \Phi_{1}<0$. Thus the sign of Jacobian $J(x_i)$, i.e., the topological charge of zero point, is finally identified by
\begin{eqnarray}
Q_i \equiv \operatorname{sgn} J(x_i)=\operatorname{sgn} (\partial_{r_+} \Phi_{2})_{x_i}.
\end{eqnarray}
According the definition of $\Phi_2$, we have $(\partial_{r_+} \Phi_{2})_{x_i}=(\partial^3_{r_+} T)_{x_i}$, and thus the topological charge is uniquely determined by the sign of $(\partial^3_{r_+} T)_{x_i}$. In fact, this is an additional condition to justify the stability at critical points (see Appendix \ref{sta_con} for more details); the critical point with positive $(\partial^3_{r_+} T)_{x_i}$, i.e., the topological charge $+1$, is thermodynamically stable, whereas the one with topological charge $-1$ is thermodynamically unstable. 

\begin{figure}
\center{\subfigure{\label{CP_-}
\includegraphics[height=4cm]{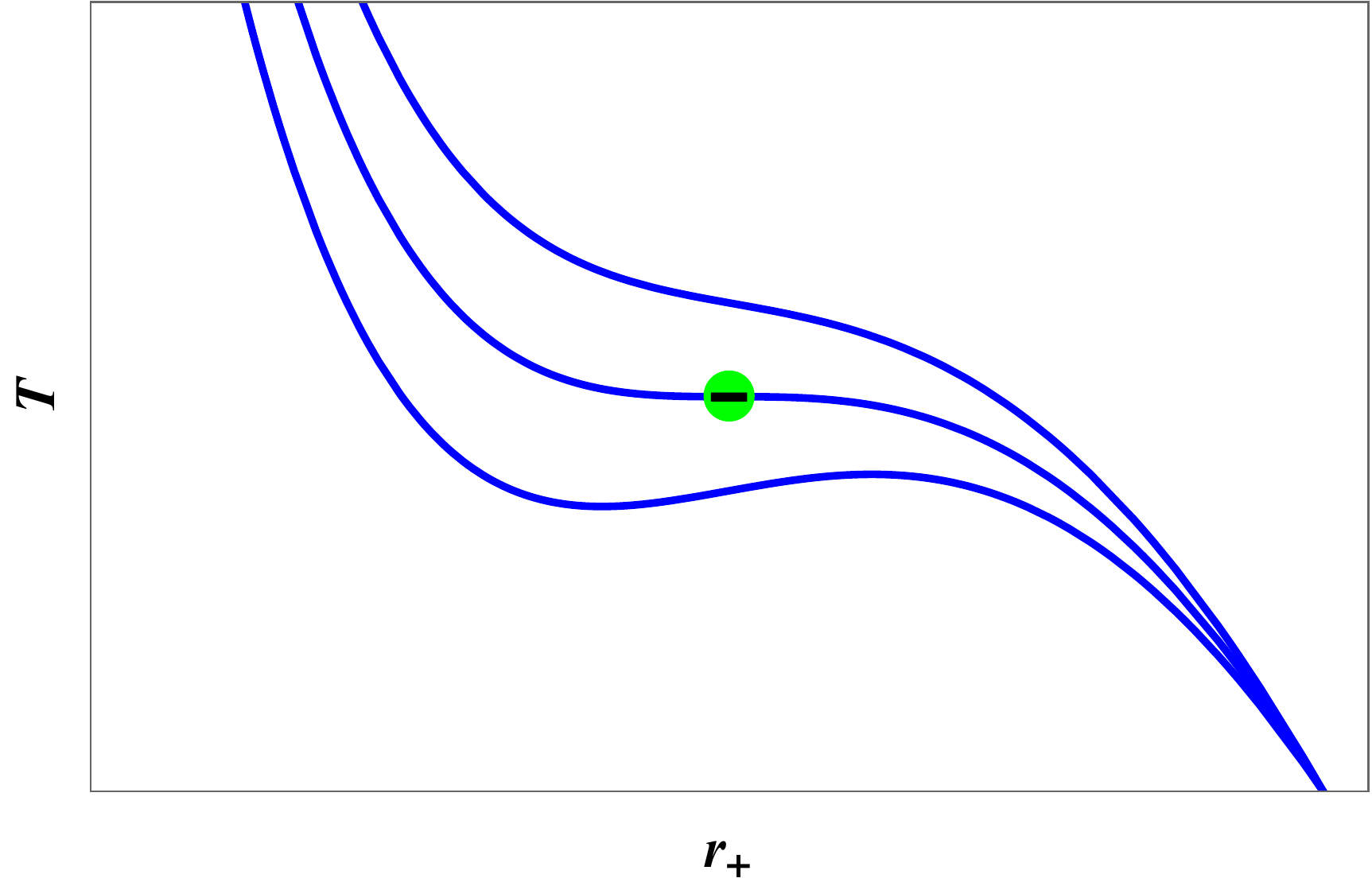}}
\hspace{1cm}
\subfigure{\label{CP_+}
\includegraphics[height=4cm]{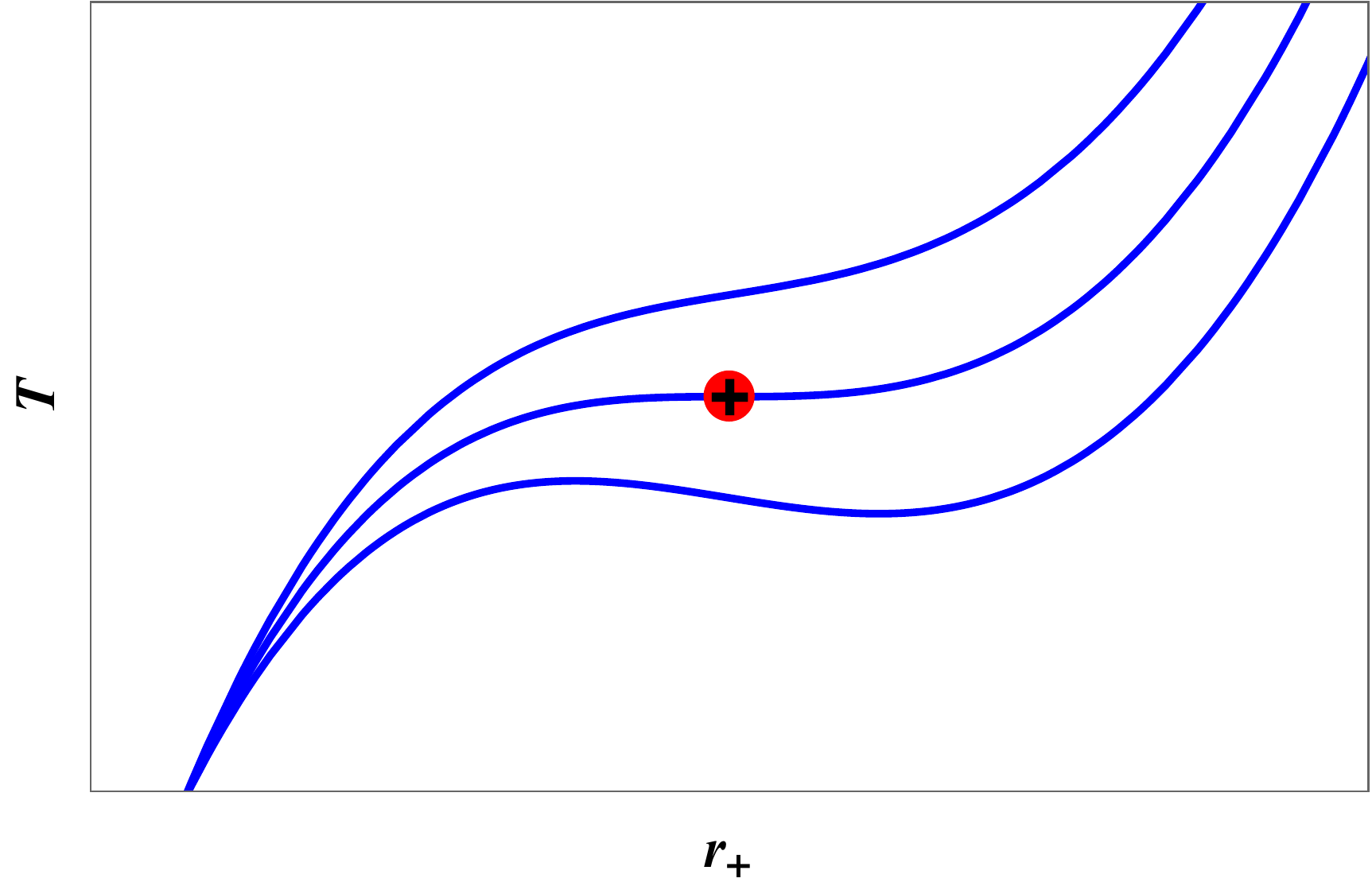}}}
\caption{\label{CP_-+} Behaviors of $T-r_+$ curves near critical points with negative (left) and positive (right) topological charges.}
\end{figure}

To be more visual, we display the sketches of $T-r_+$ curves near two kinds of critical points in Fig. \ref{CP_-+}. Near the $+1$ one, the stable black hole branches (heat capacity $C_{C}=T\left(\partial_{S} T\right)^{-1}_{C,z^i}>0$) are on two sides, while the unstable black hole branch is in the middle, and one can draw a first-order phase transition line in $T-S$ plane among these stable black hole branches using Maxwell’s equal area law. A possible second-order phase transition is expected at this kind of critical point. 

On the contrary, near the $-1$ one, the unstable black hole branches are on two sides, whereas the stable black hole branch is in the middle, and thus one can not draw a first-order phase transition line. Of course, a second-order phase transition can not occur at this kind of critical point.

For the cases shown in Fig. \ref{d=4}, $\text{CP}_2$ and $\text{CP}$ are stable critical points, at which the second-order phase transition can actually occur, while $\text{CP}_1$ is an unstable critical point at which the second-order phase transition is forbidden. In the phase diagrams, one can expect that $\text{CP}_2$ and $\text{CP}$ directly connect to the first-order phase transition line, while $\text{CP}_1$ becomes an `isolated' critical point which does not connect to the first-order phase transition line.

%%%%%%%%%%%%%%%%%%%%%%%%%%%
%%%%%%%%%%%%%%%%%%%%%%%%%%%

\section{Phase transition}\label{phase_tran}

Until now, using the global topological analysis, we have found that there is a high probability of finding a reentrant phase transition in the $D=4$ black hole, and a high probability of finding VdW phase transition or triple point in the $D>4$ black hole. The rough phase structure has also been described by performing the local topological analysis. In this section, whether these phase transitions can actually occur will be carefully examined by the free energy. 

\subsection{\textit{D}=4 case}

\begin{figure}
\center{\subfigure{\label{F_T_4}
\includegraphics[height=4.5cm]{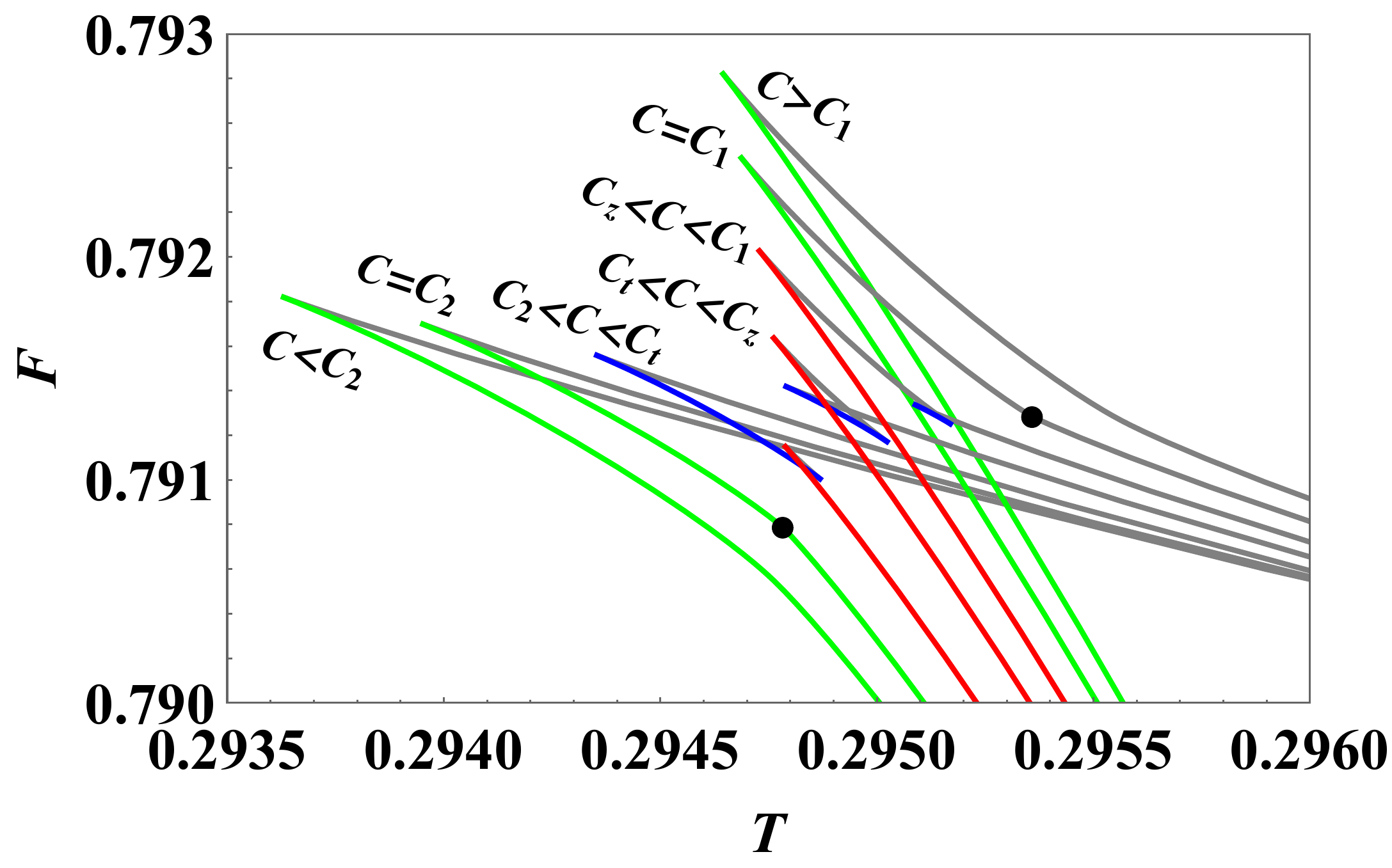}}
\hspace{1cm}
\subfigure{\label{RPT}
\includegraphics[height=4.5cm]{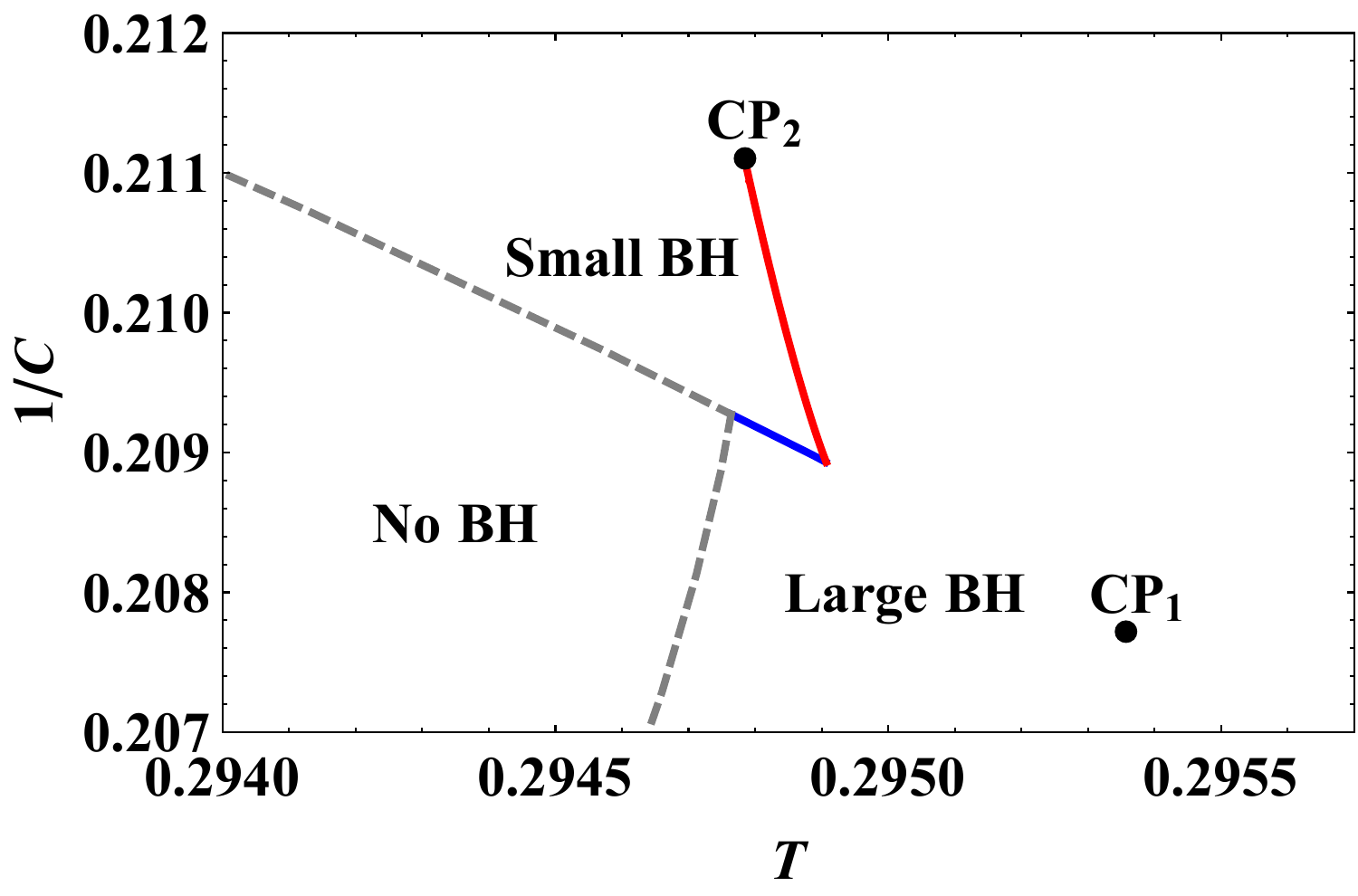}}}
\caption{\label{F_T_RPT} Left: $F-T$ diagram for $D=4$ case. Here $(C_2,C_t,C_z,C_1)=(4.737,4.778,4.786,4.814)$. Right: Reentrant phase diagram in $1/C-T$ plane. We have set $(Q,P,b)=(1,1,2.3)$.}
\end{figure}

Utilizing Eq. (\ref{eq:free}), the $F-T$ curves for different central charges at fixed $(Q,P,b,D)=(1,1,2.3,4)$ are displayed in the left panel of Fig. \ref{F_T_RPT}. Here $C_1$ and $C_2$ denote the central charges for critical points $\text{CP}_1$ and $\text{CP}_2$ shown in Fig. \ref{d=4}, respectively. The shadowed lines correspond to unstable black hole branches with negative heat capacity, whereas the colored lines represent local stable black hole branches with positive heat capacity. For $C < C_2$ or $C > C_1$, there is only one stable branch and no phase transition exists. For $C_2 < C < C_1$, two local stable black hole branches appear; the blue one has smaller $r_+$, while the red one has larger $r_+$, and we would like to refer them as small black hole (SBH) branches and large black hole (LBH) branches respectively. Specially, the first-order phase transition takes place between these branches in the range of $C_2<C<C_z$, which can be seen from the swallowtail behavior of $F-T$ curves. When $C=C_2$, the swallowtail turns into a point, at which the second-order phase transition takes place, corresponding to a stable critical point. Note that the second-order phase transition does not actually take place when $C=C_1$, for the critical point here (black dot) connects with unstable branches. This is consistent with our topological analysis.

Besides the first-order and second-order phase transitions, we also observe a zeroth-order phase transition in the range of $C_t<C<C_z$. Starting with the lowest temperature, the system evolves along the LBH branch (red) first. Increasing the temperature, there is a discontinuous global minimum in the $F-T$ curve, and thus the system jumps form the LBH branch to the SBH branch (blue)---this refers to a zeroth-order phase transition. Further increasing the temperature, the system undergoes a first-order phase transition and eventually reenters the LBH branch. These successive phase transitions exactly correspond to the reentrant phase transition we are looking for.

The phase diagram for the reentrant phase structure is shown in the $1/C-T$ plane, see the right panel of Fig. \ref{F_T_RPT}. The first-order phase transition line (red solid curve) connects with the zeroth-order phase transition line (blue solid curve), and terminates at the critical point $\text{CP}_2$. Upon these two phase transition lines, it is the region of SBHs, and below them the region of LBHs occupies. There is also a region where no black hole exists, bounded by two dashed curves which correspond to the left vertices of $F-T$ curves. Another critical point $\text{CP}_1$ is `isolated', as expected from our topological analysis.

%%%%%%%%%%%%%%%%%%%%%%%%%%%
%%%%%%%%%%%%%%%%%%%%%%%%%%%

It is worth noting that a standard VdW like phase transition is also found in the $D = 4$ case, where the $F-T$ curve only admit three branches (two stable branches and one unstable branches), unlike the case of four branches we presented above. To search for this phase behavior in our parameter setting, first notice that the number (even or odd) of black hole branches can be justified by the number of extrema in the $T-r_+$ curve, or equally, by the number of zeros of $\Phi_1=(\partial_{r_+} T)_{C,z^i}=0$. To capture this information, we can study the asymptotic behavior of $\Phi_1$:
\begin{align}
\Phi_1(r_+ \rightarrow 0^+) \sim \frac{\sqrt{3/ 2 \pi } b Q-\sqrt{P C}}{4 \pi r_+^2 \sqrt{P C}}, \quad \Phi_1(r_+ \rightarrow + \infty) \sim \sqrt{\frac{3 P}{2 \pi  C}}.
\end{align}
For $C<3 b^2 Q^2/2 \pi P=C_*$, $\Phi_1(r_+ \rightarrow 0^+)>0$, $\Phi_1(r_+ \rightarrow + \infty)>0$, and thus $\Phi_1$ admits even number of zeros, indicating the black hole has an odd number of branches. While for $C>C_*$, $\Phi_1(r_+ \rightarrow 0^+)<0$, $\Phi_1(r_+ \rightarrow + \infty)>0$, and $\Phi_1$ admits an odd number of zeros, resulting in even number of black hole branches. A detailed topological analysis shows that when $P<0.029 b^3 Q$, there will be at least one critical point with positive topological charge in the $C<C_*$ region (see Appendix \ref{VdW_point}). This explains the appearance of a standard VdW like critical point. Note that there will also exists accompanying critical points (with negative topological charge) in the $C>C_*$ region, to ensure the total topological charge $Q=0$, see Appendix \ref{VdW_point} for more details. One can expect that between the $C_*$ and the central charge for the accompanying critical point, similar reentrant phase behavior shown in the left panel of Fig. \ref{F_T_RPT} takes place. Specially, if the $b$ or $Q$ is large, and the pressure $P$ has a small value (or a large AdS radius $l$), the reentrant phase behavior can be observed in the large $C$ region, where the gravity dual to the CFT expect to be classical.

\subsection{\textit{D}>4 case}

%%%%%%%%%%%%%%%%%%%%%%%%%%%
\begin{figure}
\center{\subfigure{\label{F_T_5}
\includegraphics[height=4.5cm]{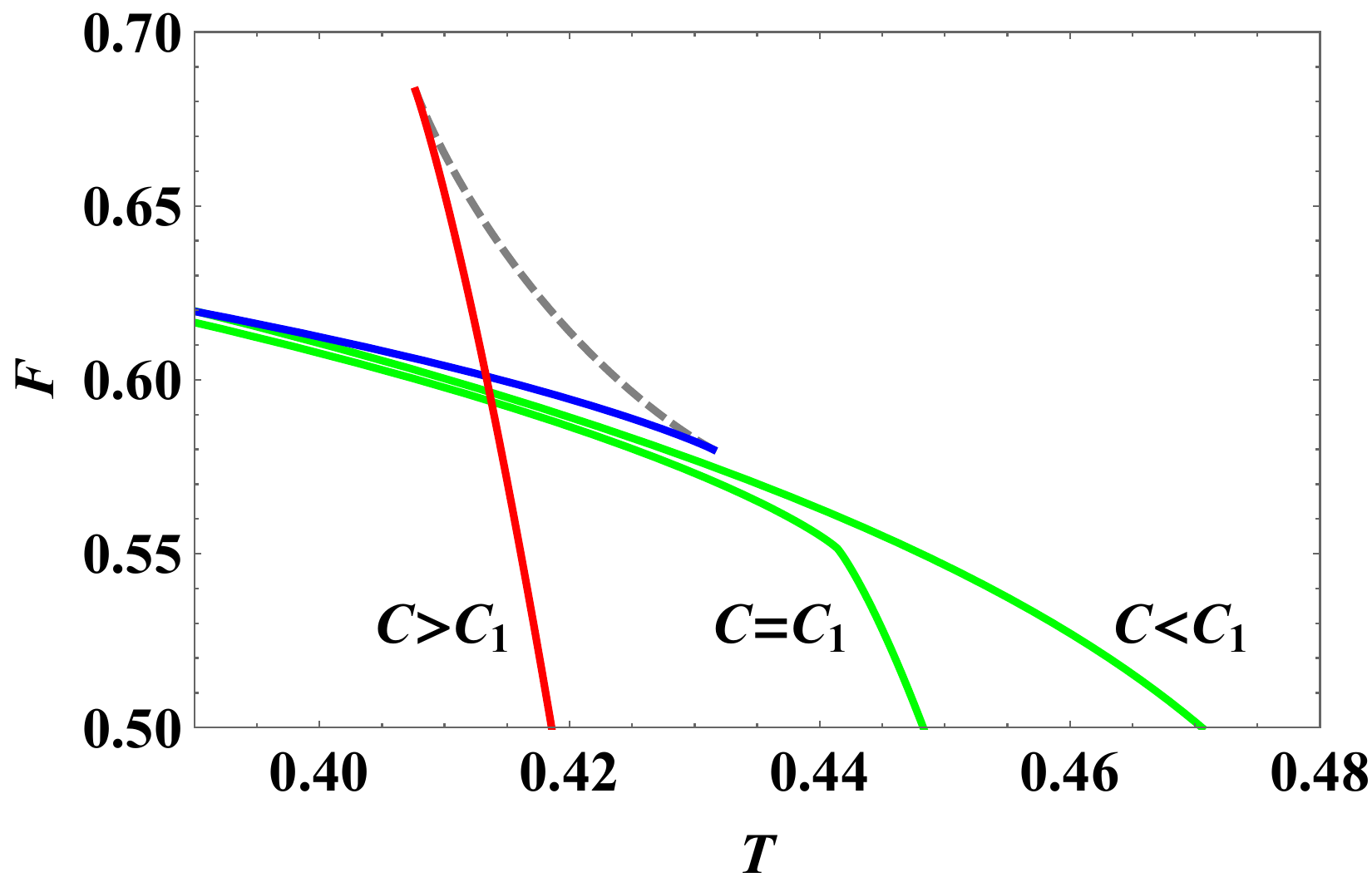}}
\hspace{1cm}
\subfigure{\label{VdW}
\includegraphics[height=4.5cm]{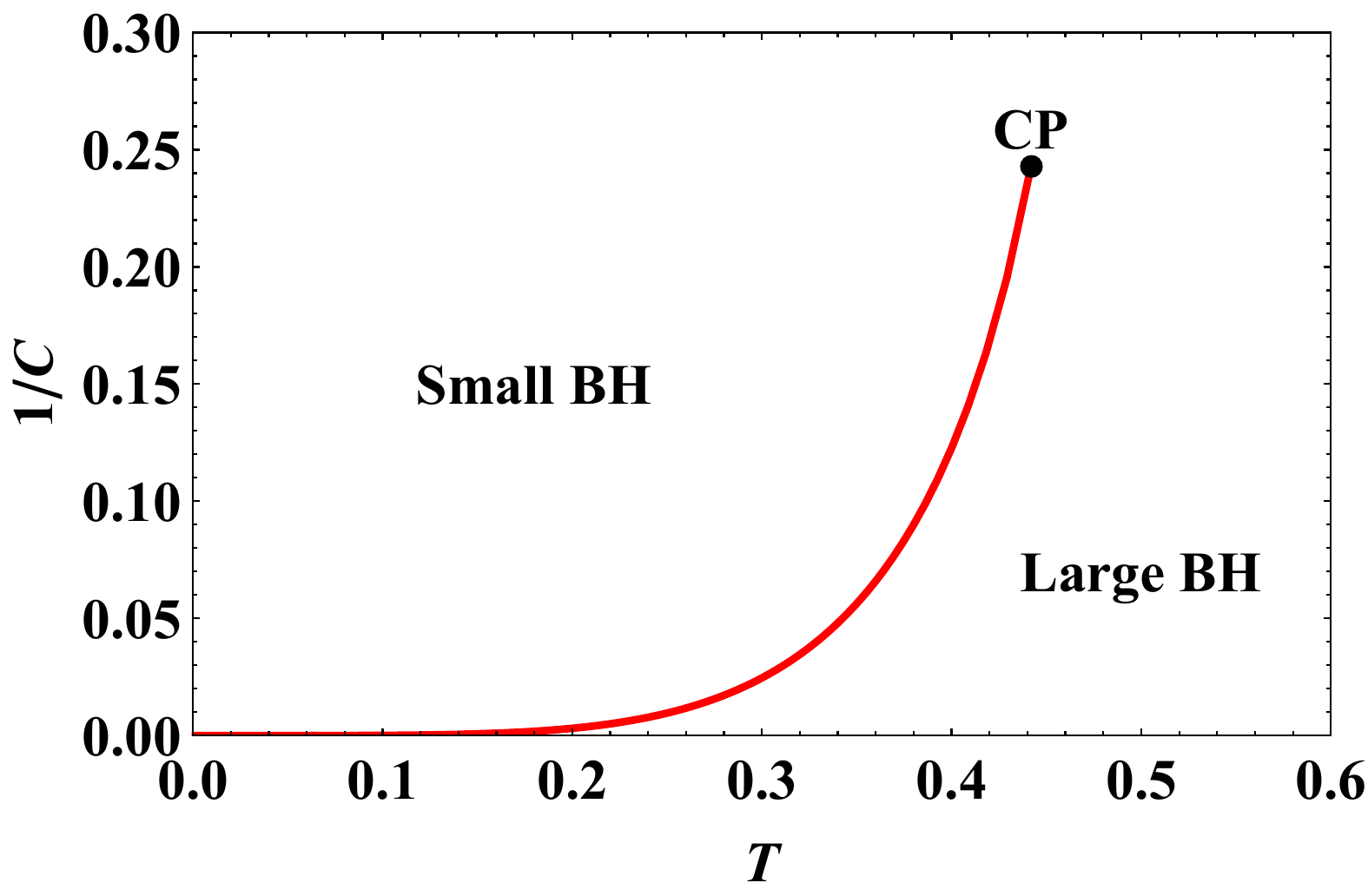}}}
\caption{\label{F_T_VdW} Left: $F-T$ diagram for $D=5$ case. The crititical central charge $C_1=4.117$. Right: VdW like phase diagram in $1/C-T$ plane. We have set $(Q,P,b)=(1,1,2.3)$.}
\end{figure}
%%%%%%%%%%%%%%%%%%%%%%%%%%%

We first examine the $D=5$ case. The $F-T$ curves for different central charges at fixed $(Q,P,b)=(1,1,2.3)$ are plotted in Fig. \ref{F_T_VdW}. Here $C_1$ denotes the central charge for critical point $\text{CP}$ shown in Fig. \ref{d=4}. When $C>C_1$, the $F-T$ curve admits swallowtail behavior, indicating a first-order phase transition between two stable branches; the blue one possessing smaller $r_+$ corresponds to SBH branch, while the red one possessing larger $r_+$ corresponds to LBH branch. When $C=C_1$, the swallowtail turns into a point, at which the second-order phase transition occurs. For $C<C_1$, the $F-T$ curve is single valued and no phase transition exists. In other words, the bulk first order phase transition can only exist for the dual CFTs with a large central charge (or many degrees of freedom). 

In the right panel of Fig. \ref{F_T_VdW}, we show the phase diagram in the $1/C-T$ plane. The first-order phase transition line, or the coexistence curve of SBHs and LBHs, beginning with the origin and terminating at the critical piont $\text{CP}$, is reminiscent of the phase transition line of VdW fluids in the $P-T$ phase diagram. This also indicates that the phase transition can appear for arbitrary small 
$1/C$, or arbitrary large $C$. This is in contrast to the reentrant case observed in $D=4$, where the phase transitions can only exist for a small range of $C$, or equally,  degrees of freedom.

We also investigate the $D=6 \sim 10$ case, in which similar VdW like phase structures are observed. The triple point is not found. For arbitrary higher-dimension, from the topological analysis, we can be sure that the phase transition behavior must exist, for at least one critical point with positive topological charge exist in the black hole. Of course, the exact type needs to be further determined by the free energy argument.

\section{Conclusions and discussions}\label{conclu}

Summarizing, we have demonstrated that the reentrant phase transition exists in the holographic thermodynamics of Born-Infeld AdS black hole. With the help of topological analysis, such a phase transition is observed in the 
$D=4$ case of fixed $(Q,P,b,C)$ ensemble in mixed bulk/boundary formalism. For higher-dimensions, the VdW like phase transition and the possible triple point replaces. 

\begin{figure}
\center{\subfigure{\label{CFT_F_T_4}
\includegraphics[height=4.5cm]{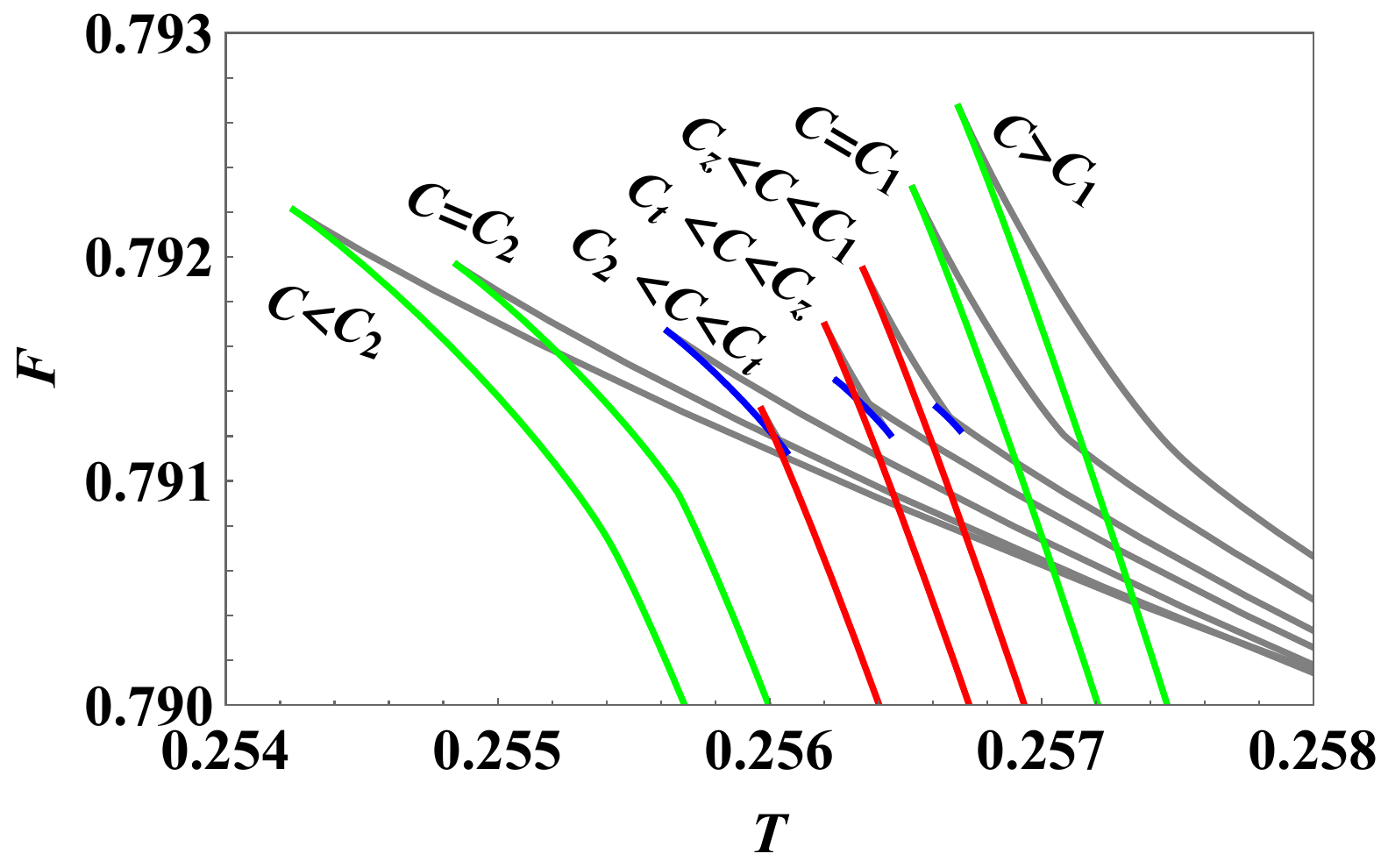}}
\hspace{1cm}
\subfigure{\label{CFT_RPT}
\includegraphics[height=4.5cm]{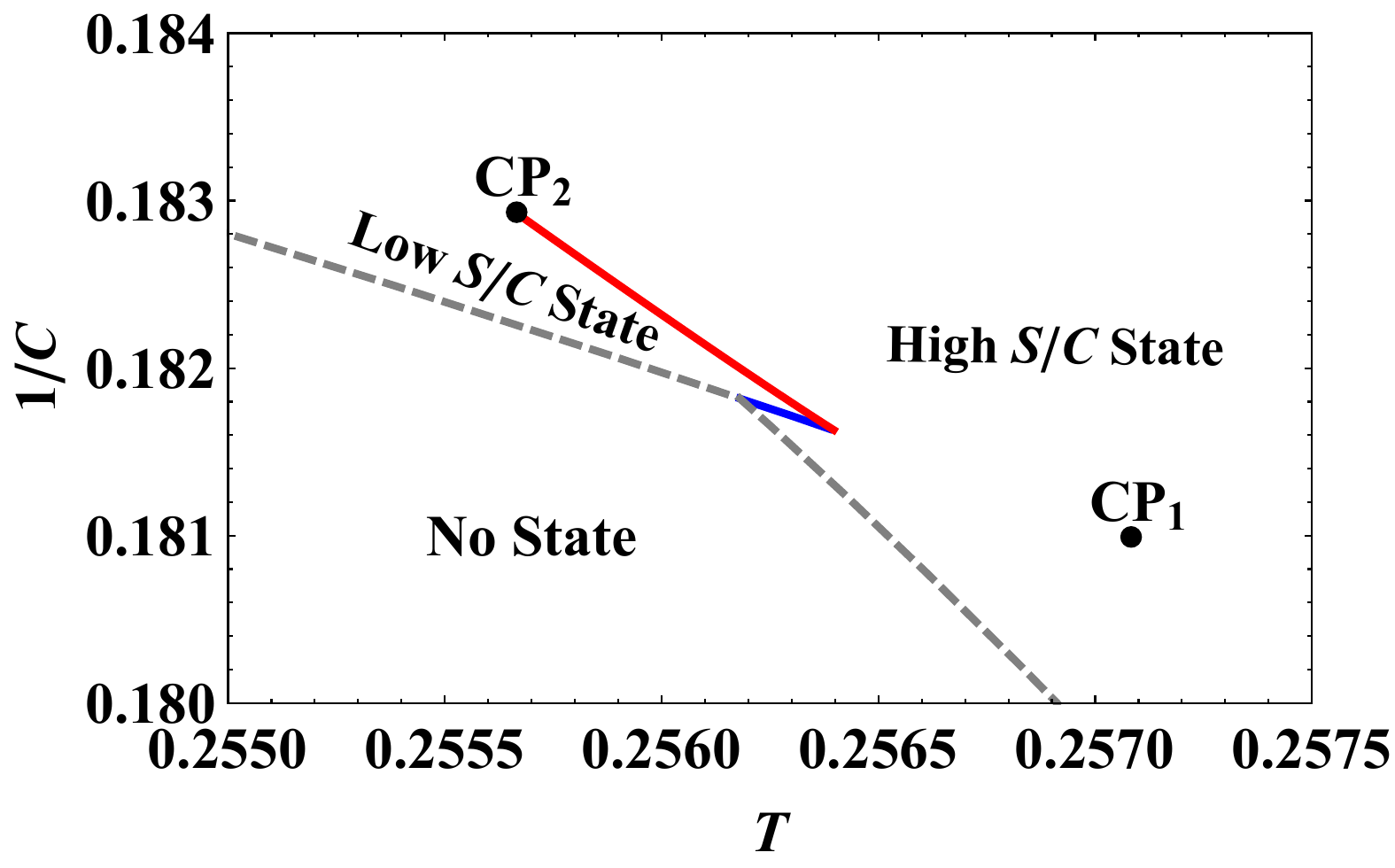}}}
\caption{\label{CFT_F_T_RPT} Left: $F-T$ diagram for CFT in $D=4$ (bulk dimension) case. Here $(C_2,C_t,C_z,C_1)=(5.467,5.5,5.506,5.525)$. Right: Reentrant phase diagram for CFT in $1/C-T$ plane. We have set $(\tilde{Q},R,\tilde{b})=(1,1,2)$.}
\end{figure}

\begin{figure}[b]
\center{\subfigure{\label{CFT_F_T_5}
\includegraphics[height=4.5cm]{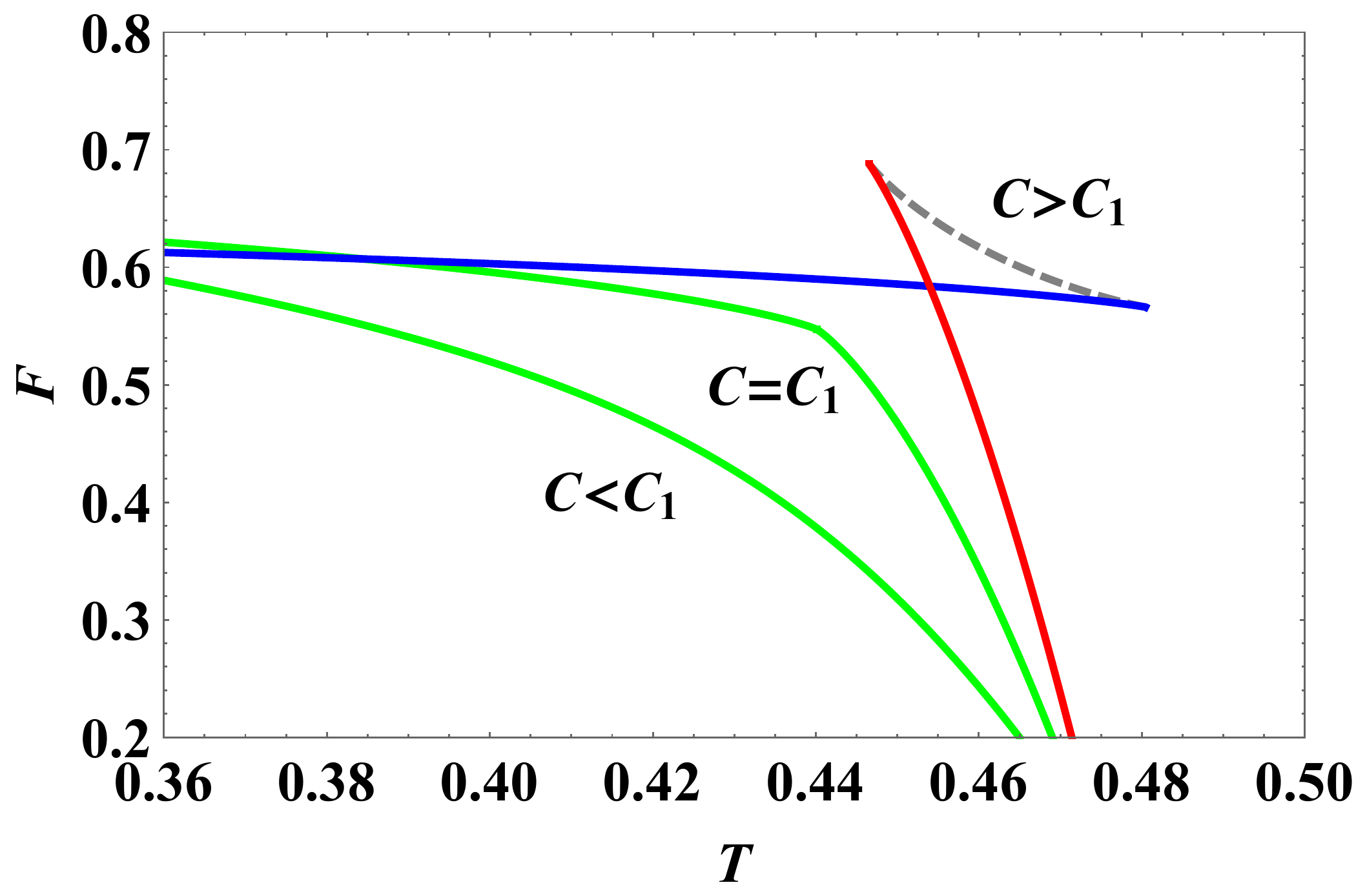}}
\hspace{1.3cm}
\subfigure{\label{CFT_VdW}
\includegraphics[height=4.5cm]{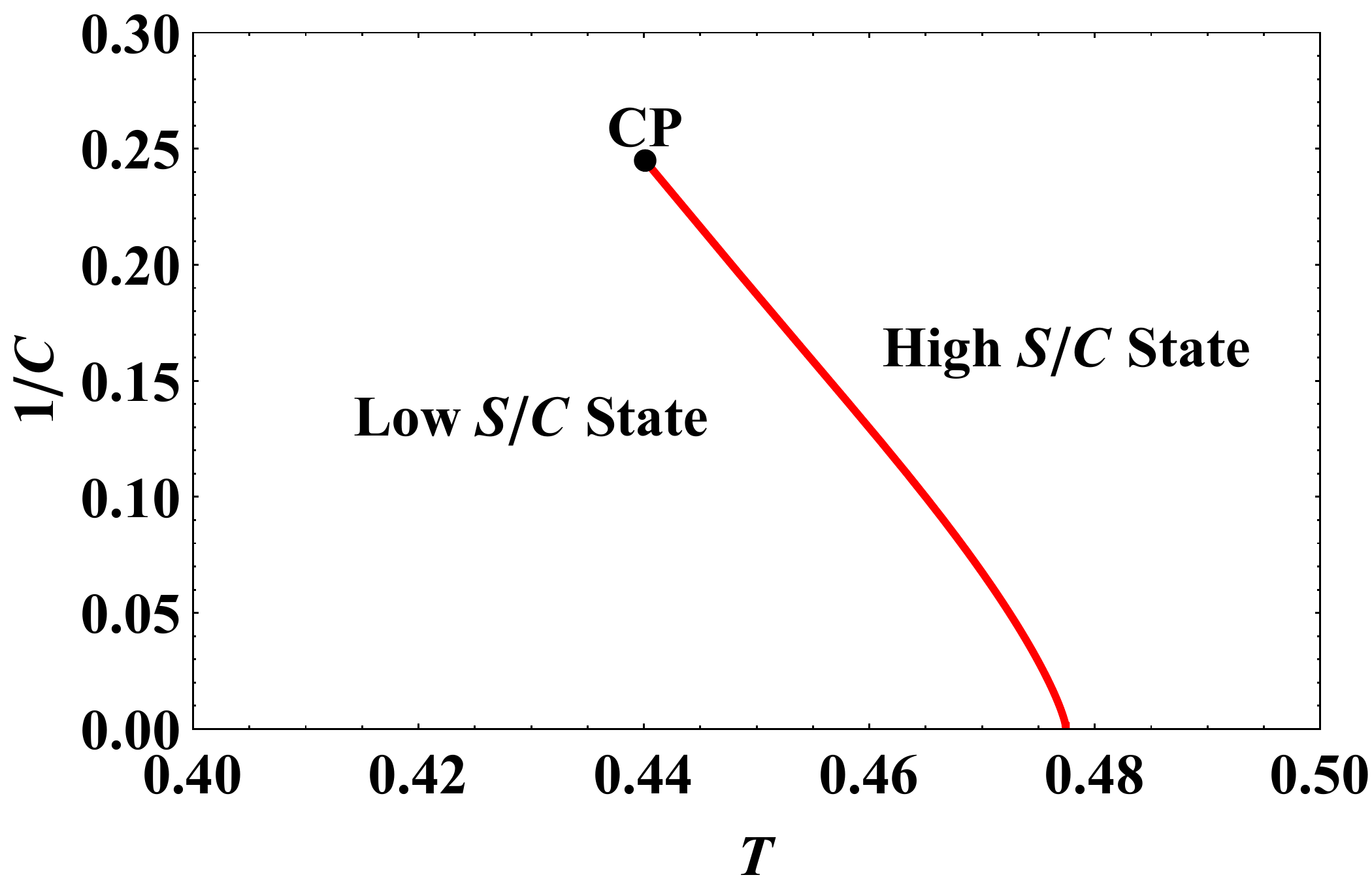}}}
\caption{\label{CFT_F_T_VdW} Left: $F-T$ diagram for CFT in $D=5$ (bulk dimension) case. The critical central charge $C_1=4.085$. Right: VdW like phase diagram for CFT in $1/C-T$ plane. We have set $(\tilde{Q},R,\tilde{b})=(1,1,2)$.}
\end{figure}

It is worth emphasizing that all the critical behaviors presented above take place in the bulk. A natural question is then whether similar critical behaviors exist in the dual CFT. To actually see this, we first introduce a dimensionless parameter $\tilde{r}=r_+/l$ suggested in Refs. \cite{Dolan:2016jjc,Cong:2021jgb}, which has the physical interpretation
of $S/C$ (up to a index $D-2$), namely the entropy per degree of
freedom for the CFT thermal states. Utilizing this parameter and the holographic dictionary (\ref{eq:dic_redefined}), the CFT temperature can be cast into
\begin{align}
T=\frac{1}{4 \pi \tilde{r} R} \Bigl[(D-3) +(D-1) \tilde{r}^2 + \frac{4 \tilde{b}^2 \tilde{r}^2 \left(1-x\right)}{D-2}\Bigr], \label{eq:tem_CFT}
\end{align}
where
\begin{align}
x=\sqrt{1+\frac{16 \pi ^2 \tilde{r}^{4-2 D} \tilde{Q}^2}{\tilde{b}^2 \omega^2 C^2}}. \label{eq:x_CFT}
\end{align}
The free energy in the fixed $(\tilde{Q},\mathcal{V},\tilde{b},C)$ ensemble, corresponding to the fixed $(Q,P,b,C)$ ensemble discussed before, can also be calculated as $ F = E-TS$, i.e.,
\begin{align}
 F &= \frac{\omega x^{D-3} C}{16 \pi  R} \Bigl[1-\tilde{r}^2+\frac{4 \tilde{b}^2 \tilde{r}^2 \left(x-1\right)}{(D-1)(D-2)}\Bigr] \nonumber \\
&\quad +\frac{4 \pi  (D-2) \tilde{r}^{3-D} \tilde{Q}^2 }{(D-1) (D-3) \omega R C} \times{ }_2 F_1\Bigl[\frac{D-3}{2 D-4},\frac{1}{2},\frac{3 D-7}{2 D-4},-\frac{16 \pi ^2 \tilde{r}^{4-2 D} \tilde{Q}^2}{\tilde{b}^2 \omega^2 C^2}\Bigr]. \label{eq:free_CFT}
\end{align}
Note that fixing the CFT volume $\mathcal{V}$ is identical to fixing the boundary curvature radius $R$. The $F-T$ curve can be then numerically plotted from Eqs. (\ref{eq:tem_CFT}--\ref{eq:free_CFT}) using $\tilde{r}$ as parameter. In the $D=4$ (bulk dimension) case, taking $(\tilde{Q},R,\tilde{b})=(1,1,2)$ for example, the $F-T$ curves for different $C$ together with the $1/C-T$ phase diagram are shown in Fig. \ref{CFT_F_T_RPT}. Similar reentrant phase behavior is observed. However, instead of the large-small-large black hole phase transition, the phase transition here takes place between the dual CFT thermal states with high-, low- and high-entropy per degree of freedom. In the $D=5$ case, we observe a VdW like phase transition between the dual CFT thermal states with high- and low-entropy per degree of freedom, as displayed in Fig. \ref{CFT_F_T_VdW}. The first-order phase transition line here has negative slope and does not start from the origin, which is different from the bulk one shown in Fig. \ref{F_T_VdW}. Even so, the phase transition can only appear for large $C$ (degrees of freedom), and interestingly, can exist for arbitrary large $C$, in contrast to the $D=4$ case. For $D=6 \sim 10$ case, similar VdW like phase transition is observed. In the future work, it would be interesting to examine whether the triple point can exist for higher-dimensions, both in the bulk and the dual CFT. Moreover, the presented topological approach for searching possible phase transitions and phase structures is also worth extending to other black holes.

%%%%%%%%%%%%%%%%%%%%%%%%%%%
%%%%%%%%%%%%%%%%%%%%%%%%%%%

%%%%%%%%%%%%%%%%%%%%%%%%%%%
%%%%%%%%%%%%%%%%%%%%%%%%%%%

\begin{acknowledgments}
We are grateful to Jin-Tao Luo for useful discussions. The authors contribute equally to this work. This work is supported by Sichuan Science and Technology Program, NO:2022YFG0317 and  NSFC (Grant No.
12275183, 12275184, 12105191, and 12175212).
\end{acknowledgments}

\appendix

\section{Calculation of central charge \textit{C}} \label{central}

We now review the calculation of central charge to determine the numerical factor $k$ in Eq. (\ref{eq:central}). The AdS/CFT correspondence \cite{Maldacena:1997re,Gubser:1998bc,Witten:1998qj} plays a pivotal role in the calculation, which can be precisely formulated by equating the partition functions of two systems:
\begin{equation}
Z_{A d S}\left(\phi_{0, i}\right)=Z_{C F T}\left(\phi_{0, i}\right). \label{eq:ads/cft}   
\end{equation}
On the gravity side, the fields $\phi_{0, i}$ correspond to the boundary values of the bulk fields $\phi_{i}$ propagating in the AdS space. On the field theory side, these fields correspond to external source currents coupled to various CFT operators $\mathcal{O}^{i}$. In a saddle–point
approximation, one can further formulate Eq. (\ref{eq:ads/cft}) as
\begin{equation}
e^{-I_{A d S}\left(\phi_{i}\right)}=\left\langle e^{\int \phi_{0, i} \mathcal{O}^{i}}\right\rangle_{C F T},   
\end{equation}
where $I_{A d S}$ is the classical gravitational action. This correspondence
enables us to evaluate the expectation values of the CFT operators in terms of the asymptotic gravity fields. Specially, the expectation value of the stress tensor of the dual field theory $\left\langle T^{\mu \nu}\right\rangle$, can be evaluated by the the quasilocal stress tensor for AdS gravity $\tau^{\mu \nu}$, defined locally on the boundary of AdS spacetime \cite{Brown:1992br,Balasubramanian:1999re},
\begin{equation}
\left\langle T^{\mu \nu}\right\rangle=\tau^{\mu \nu}=\frac{2}{\sqrt{-\gamma}} \frac{\delta I_{AdS}}{\delta \gamma_{\mu \nu}}, \label{eq:quasilocal}  
\end{equation}
where $\gamma_{\mu \nu}$ is the boundary metric. Known as Weyl anomaly \cite{Duff:1993wm}, the trace of $\left\langle T^{\mu \nu}\right\rangle$ vanishes in 
flat spacetime, but can have a non-zero valve in curved spacetime. For example, the $2$-dimensional CFT has a anomaly
\begin{equation}
\left\langle T^{\mu}_{\mu}\right\rangle = - \frac{C}{24 \pi} \mathcal{R}, \label{eq:anomaly} 
\end{equation}
where $C$ is the central charge of the CFT, and $\mathcal{R}$ is the curvature of the background spacetime. With Eqs. (\ref{eq:quasilocal}) and (\ref{eq:anomaly}), one can relate the central charge of the CFT to the parameters in gravity theory. For Einstein gravity, the action $I_{AdS}$ for $D$-dimensional gravity can be generally expressed as \cite{Balasubramanian:1999re,Emparan:1999pm}
\begin{equation}
I_{A d S}=-\frac{1}{16 \pi G} \int_{\mathcal{M}} d^{D} x \sqrt{-g}\bigl( R-2 \Lambda \bigr) - \frac{1}{8 \pi G} \int_{\partial \mathcal{M}} d^{D-1} x \sqrt{-\gamma} \Theta + I_{matter} + I_{ct}.   
\end{equation}
The second term corresponds to the boundary term, which is required to obtain the correct field equations. The last term corresponds to the counterterm that is added to obtain a finite gravitational action. $\Theta$ is the trace of the extrinsic curvature of the boundary, defined as $\Theta=\gamma^{i j} \Delta_{i} n_{j}$ ($n_{j}$ the outward pointing unit normal vector to $\partial \mathcal{M}$). Note that since we always consider solutions to the field equations, i.e. the on-shell condition, only the boundary term and counterterm contributes to the variation of action. In other words, the type of matter field will not affect the calculation of $\left\langle T^{\mu \nu}\right\rangle$ or $C$.

Choosing the coordinates
\begin{equation}
d s^{2}=\frac{l^2}{r^2} d r^{2} + \gamma_{\mu \nu} d x^{\mu} d x^{\nu},  \label{eq:metric} 
\end{equation}
and further taking the $r \rightarrow \infty$ limit, to the leading order, one can obtain the exact expression for $\tau^{\mu}_{\mu}$ in $3$-dimensional gravity \cite{Balasubramanian:1999re,Henningson:1998gx}
\begin{equation}
\tau^{\mu}_{\mu} = - \frac{l}{16 \pi G} \mathcal{R}.  
\end{equation}
where $\mathcal{R}$ is the curvature of the metric $\gamma_{\mu \nu}$. Comparing this with Eq. (\ref{eq:anomaly}), we can find
\begin{equation}
C = \frac{3 l}{2 G}.  
\end{equation}
This agrees with the value of $C$ as computed in Ref. \cite{Brown:1986nw} by considering the asymptotic symmetry algebra of $\text{AdS}_3$.

Following the similar procedure, we can read off the 
central charge in higher dimensions. In $5$-dimension, the AdS gravity is expected to be dual to $4$-dimensional $\mathcal{N}=4$, $SU(N)$ super Yang-Mills theory. In the large $N$ limit, the anomaly of the Yang-Mills theory is given by \cite{Duff:1993wm}
\begin{equation}
\left\langle T^{\mu}_{\mu}\right\rangle=-\frac{N^{2}}{4 \pi^2}\left[-\frac{1}{8} \mathcal{R}^{\mu \nu} \mathcal{R}_{\mu \nu}+\frac{1}{24} \mathcal{R}^{2}\right].  \label{eq:Yang}
\end{equation}
On the gravity side, considering the coordinates Eq. (\ref{eq:metric}) and the $r \rightarrow \infty$ limit, one can find \cite{Balasubramanian:1999re,Henningson:1998gx}
\begin{equation}
\tau^{\mu}_{\mu}=-\frac{l^{3}}{8 \pi G}\left[-\frac{1}{8} \mathcal{R}^{\mu \nu} \mathcal{R}_{\mu \nu}+\frac{1}{24} \mathcal{R}^{2}\right].  
\end{equation}
Comparing this with Eq. (\ref{eq:Yang}), we can read off
\begin{equation}
C=N^2=\frac{\pi l^3}{2 G}.  
\end{equation}
For generic number
of dimensions, the central charge can be expected to be
\begin{equation}
C = k \frac{l^{D-2}}{G},  
\end{equation}
where the numerical factor $k$ is determined by the dimension, and as discussed above, remains unaffected by the type of matter field.

\section{Topological charge for \textit{D}>4 case} \label{higher_d}

For $D > 4$ case, the components of vector field $\Phi$ have been given in Eqs. (\ref{eq:first_con}) and (\ref{eq:second_con}). Similarly, we now examine the behavior of vector field $\Phi$ on these line segments shown in Fig. \ref{fig:contour}. Along $l_{1}$, we have $r_+=\delta^{-1}$, $C=\delta^{-\beta}$, and
\begin{equation}
x (\delta \rightarrow 0^+) \sim 1 + \frac{16^{\frac{4}{D}-1} \pi ^{4/D} Q^2 y^{2-\frac{4}{D}}}{2 b^2 \omega ^2} \delta ^{4 \beta/D+2 D-4}
\end{equation}
for $\beta \in [-1,1]$, which results in
\begin{align}
&\Phi_1 (\delta \rightarrow 0^+) \sim \frac{(D-1) (16 \pi )^{2/D} y^{-2/D} }{4 \pi } \delta ^{2 \beta/D},\\
&\Phi_2 (\delta \rightarrow 0^+) \sim \frac{D-3}{2 \pi } \delta ^3.
\end{align}
Since $|\Phi_2/\Phi_1| \rightarrow 0$ and $\Phi_1 > 0$, the vector 
$\Phi$ is horizontal to the right at $l_1$, i.e., $\Omega_{l_1}=0$, indicating that
\begin{eqnarray}
 \Delta\Omega_{l_1}=0.
\end{eqnarray}
\noindent Along $l_{2}$, we have $r_+=\delta^{-\alpha}$, $C=\delta^{-1}$, and
\begin{align}
 x (\delta \rightarrow 0^+) \sim \left\{\begin{array}{ll}
\frac{16^{\frac{2}{D}-\frac{1}{2}} \pi ^{2/D} y^{1-\frac{2}{D}} Q}{b \omega } \delta ^{(D-2) \alpha +\frac{2}{D}} & \text { for } \alpha \in [-1, - \frac{2}{D (D-2)}), \\
\left[1+\frac{16^{\frac{4}{D}-1} \pi ^{4/D} y^{2-\frac{4}{D}} Q^2}{b^2 \omega ^2} \right]^{1/2} & \text { for } \alpha = - \frac{2}{D (D-2)}, \\
1+\frac{16^{\frac{4}{D}-1} \pi ^{4/D} y^{2-\frac{4}{D}} Q^2}{2 b^2 \omega ^2} \delta ^{(2 D-4) \alpha +\frac{4}{D}} & \text { for } \alpha \in (- \frac{2}{D (D-2)}, 1],
\end{array}\right.
\end{align}
which gives
\begin{align}
& \Phi_1 (\delta \rightarrow 0^+) \sim \left\{\begin{array}{ll}
\frac{(D-3) 2^{8/D} \pi ^{2/D} y^{1-\frac{2}{D}} b Q}{4 \pi (D-2) \omega } \delta ^{(D-2) \alpha+\frac{2}{D}} & \text { for } \alpha \in [-1, - \frac{2}{D (D-4)}), \\
\left[\frac{2^{8/D} (D-3) \pi ^{2/D} y^{1-\frac{2}{D}} b Q}{4 \pi  (D-2) \omega }-\frac{D-3}{4 \pi }\right] \delta ^{-\frac{4}{D (D-4)}} & \text { for } \alpha = - \frac{2}{D (D-4)}, \\
-\frac{D-3}{4 \pi } \delta ^{2 \alpha } & \text { for } \alpha \in (- \frac{2}{D (D-4)}, 1/D), \\
\frac{2^{8/D} (D-1) \pi ^{2/D} y^{-2/D}-(D-3)}{4 \pi } \delta ^{2/D} & \text { for } \alpha = 1/D, \\
\frac{2^{8/D} (D-1) \pi ^{2/D} y^{-2/D}}{4 \pi } \delta ^{2/D} & \text { for } \alpha \in (1/D, 1],
\end{array}\right. \\
& \Phi_2 (\delta \rightarrow 0^+) \sim \left\{\begin{array}{ll}
-\frac{2^{8/D} (d-3) \pi ^{2/D} y^{1-\frac{2}{D}} b Q}{4 \pi  \omega } \delta ^{(D-1) \alpha +\frac{2}{D}} & \text { for } \alpha \in [-1, - \frac{2}{D (D-4)}), \\
\left[\frac{D-3}{2 \pi }-\frac{2^{8/D} (D-3) \pi ^{2/D} y^{1-\frac{2}{D}} b Q}{4 \pi  \omega }\right] \delta ^{-\frac{3}{D (D-4)}} & \text { for } \alpha = - \frac{2}{D (D-4)}, \\
\frac{D-3}{2 \pi } \delta ^{3 \alpha } & \text { for } \alpha \in (- \frac{2}{D (D-4)}, 1],
\end{array}\right.
\end{align}
Since $|\Phi_2/\Phi_1| \rightarrow 0$ and $\Phi_1 > 0$ when $\alpha=1$, the vector field $\Phi$ is horizontal to the right at the joint of $l_1$ and $l_2$, as expected from our calculation on $l_1$. At the joint of $l_2$ and $l_3$ ($\alpha=-1$), $|\Phi_2/\Phi_1| \rightarrow + \infty$ and $\Phi_2 < 0$, thus the vector field $\Phi$ points down vertically and $\Omega$ may changes $3 \pi/2 + 2 k \pi$ in anti-clockwise direction or $\pi/2 + 2 k \pi$ in clockwise direction, where $k \in (0, 1, 2, \cdots)$. Since $\Phi_2$ crosses the zero point only once, we know that $k=0$. Moreover, since there exists a range for $\alpha \in (- \frac{2}{D (D-4)}, 1/D)$ where $\Phi_1<0$ and $\Phi_2>0$, we can justify that $\Omega$ changes $3 \pi/2$ in anti-clockwise direction along $l_2$, i.e.,
\begin{eqnarray}
\Delta\Omega_{l_2} &=& \frac{3 \pi}{2}.
\end{eqnarray}
Along $l_{3}$, we have $r_+=\delta$, $C=\delta^{-\beta}$, and
\begin{align}
x (\delta \rightarrow 0^+) \sim \frac{16^{\frac{2}{D}-\frac{1}{2}} \pi ^{2/D} y^{1-\frac{2}{D}} Q}{b \omega } \delta^{-(D-2) + 2 \beta/D}
\end{align}
for $\beta \in [-1, 1]$, which indicates that
\begin{align}
&\Phi_1 (\delta \rightarrow 0^+) \sim \frac{2^{8/D} (D-3) \pi ^{2/D} y^{1-\frac{2}{D}} b Q}{4 \pi  (D-2) \omega } \delta ^{-(D-2) + 2 \beta/D},\\
&\Phi_2 (\delta \rightarrow 0^+) \sim -\frac{2^{8/D} (D-3) \pi ^{2/D} y^{1-\frac{2}{D}} b Q}{4 \pi  \omega } \delta ^{-(D-1)+2 \beta/D}.
\end{align}
Since $|\Phi_2/\Phi_1| \rightarrow + \infty$ and $\Phi_2$ is negative, the vector $\Phi$ points down vertically on $l_{3}$, suggesting that
\begin{eqnarray}
\Delta\Omega_{l_3}=0.
\end{eqnarray}
Along $l_{4}$, we have $r_+=\delta^{-\alpha}$, $C=\delta$, and
\begin{eqnarray}
 x (\delta \rightarrow 0^+) \sim \left\{\begin{array}{ll}
\frac{16^{\frac{2}{D}-\frac{1}{2}} \pi ^{2/D} y^{1-\frac{2}{D}} Q}{b \omega } \delta^{(D-2) \alpha - 2/D} & \text { for } \alpha \in [-1, \frac{2}{D (D-2)}), \\
\left[1+\frac{16^{\frac{4}{D}-1} \pi ^{4/D} y^{2-\frac{4}{D}} Q^2}{b^2 \omega ^2}\right]^{1/2} & \text { for } \alpha = \frac{2}{D (D-2)}, \\
1 + \frac{16^{\frac{4}{D}-1} \pi ^{4/D} y^{2-\frac{4}{D}} Q^2}{2 b^2 \omega ^2} \delta ^{(2 D -4) \alpha - 4/D} & \text { for } \alpha \in (\frac{2}{D (D-2)}, 1],
\end{array}\right.
\end{eqnarray}
which results in
\begin{align}
& \Phi_1 (\delta \rightarrow 0^+) \sim \left\{\begin{array}{ll}
\frac{ 2^{8/D} (D-3) \pi ^{2/D} y^{1-\frac{2}{D}} b Q}{4 \pi  (D-2) \omega } \delta ^{(D-2) \alpha-\frac{2}{D}} & \text { for } \alpha \in [-1, 0), \\
\left[\frac{2^{8/D} (D-3) \pi ^{2/D} y^{1-\frac{2}{D}} b Q}{4 \pi  (D-2) \omega }+\frac{(D-1) (16 \pi )^{2/D} y^{-2/D}}{4 \pi }\right] \delta ^{-\frac{2}{D}} & \text { for } \alpha = 0, \\
\frac{(D-1) (16 \pi )^{2/D} y^{-2/D}}{4 \pi } \delta ^{-\frac{2}{D}} & \text { for } \alpha \in (0, 1],
\end{array}\right. \\
& \Phi_2 (\delta \rightarrow 0^+) \sim \left\{\begin{array}{ll}
-\frac{2^{8/D} (D-3) \pi ^{2/D} y^{\frac{d-2}{D}} b Q}{4 \pi  \omega } \delta ^{(D-1) \alpha -\frac{2}{D}} & \text { for } \alpha \in [-1, \frac{2}{D (D-2)}), \\
-\frac{16^{1+\frac{4}{D}} (2 D-5) b^2 \omega ^2 y^2 Q^2 +2^{32/D} (D-3) \pi ^{4/D} y^{4-\frac{4}{D}} Q^4}{\sqrt{1+\frac{16^{\frac{4}{D}-1} \pi ^{4/D} Q^2 y^{2-\frac{4}{D}}}{b^2 \omega ^2}} \left(16^{1+\frac{4}{D}} \pi \omega ^2 y^2 Q^2 + 2^8 \pi^{1-\frac{4}{D} }  b^2 \omega ^4 y^{4/D}\right)} \delta ^{\frac{2}{D (D-2)}} & \text { for } \alpha = \frac{2}{D (D-2)}, \\
-\frac{(2 D-5) (16 \pi )^{\frac{4}{D}-1} y^{2-\frac{4}{D}} Q^2}{\omega ^2} \delta ^{(2 D-3) \alpha -\frac{4}{D}} & \text { for } \alpha \in (\frac{2}{D (D-2)}, \frac{2}{D (D-3)}), \\
\left[\frac{D-3}{2 \pi }-\frac{(2 D-5) (16 \pi )^{\frac{4}{D}-1} y^{2-\frac{4}{D}} Q^2}{\omega ^2}\right] \delta ^{\frac{6}{D (D-3)}} & \text { for } \alpha = \frac{2}{D (D-3)}, \\
\frac{D-3}{2 \pi } \delta ^{3 \alpha } & \text { for } \alpha \in (\frac{2}{D (D-3)}, 1], 
\end{array}\right.
\end{align}
Varying $r_+$ from $0$ to $\infty$ (or $\alpha$ from $-1$ to $1$), $\Phi_1$ is always positive whereas $\Phi_2$ varies from negative to positive, and $|\Phi_2/\Phi_1|$ changes from $+ \infty$ to $0$, suggesting that $\Omega$ changes $\pi/2$ in anti-clockwise direction along $l_4$, i.e.,
\begin{eqnarray}
\Delta\Omega_{l_4}=\pi/2.
\end{eqnarray}

Then the total topological charge for $D>4$ case can be identified as
\begin{eqnarray}
Q=\frac{1}{2 \pi} \sum_{i} \Delta \Omega_{l_i}=+1.
\end{eqnarray}

\section{Stability condition for critical points} \label{sta_con}

We now derive the stability condition for critical points in the mixed bulk/boundary formalism, by use of
a simple fluctuation analysis following from Refs. \cite{chandler1988introduction,chimowitz2005introduction}. From the mixed first law (\ref{eq:first_mixed}), for a black hole system in stable equilibrium, the following inequality should be satisfied:
\begin{align}
\left(\Delta M\right)_{S,C,z^i}=\left(\delta M\right)_{S,C,z^i}+\frac{1}{2 !} \left(\delta^2 M\right)_{S,C,z^i}+\frac{1}{3 !} \left(\delta^3 M\right)_{S,C,z^i}+\frac{1}{4 !} \left(\delta^4 M\right)_{S,C,z^i}+\cdots > 0.
\end{align}
Consider the following partitioning of the system into
two independent subsystems:
\begin{equation}
\delta S=0=\delta S^{(1)}+\delta S^{(2)}.
\end{equation}
For the first-order term, one can get
\begin{equation}
\left(\delta M\right)_{S,C,z^i}=\delta S^{(1)} \left[\left(\frac{\partial M}{\partial S}\right)_{C,z^i}^{(1)}-\left(\frac{\partial M}{\partial S}\right)_{C,z^i}^{(2)}\right] \geq 0.
\end{equation}
Since this equation must hold both for $\delta S^{(1)}$ positive and negative, the term in the square brackets above must be zero to satisfy this inequality, and thus $\left(\delta M\right)_{S,C,z^i}=0$. Moreover, from Eq. (\ref{eq:first_mixed}), it follows that $\left(\partial_S M\right)_{C,z^i}=T$, and thus $T^{(1)}=T^{(2)}=T$.
For the second-order term, one can get
\begin{equation}
\left(\delta^{2} M\right)_{S,C,z^i}=\left(\delta S^{(1)}\right)^{2}\left[\left(\frac{\partial T}{\partial S}\right)_{C,z^i}^{(1)}+\left(\frac{\partial T}{\partial S}\right)_{C,z^i}^{(2)}\right].
\end{equation}
If $\left(\Delta M\right)_{S,C,z^i}>0$ and $\left(\delta M\right)_{S,C,z^i}=0$, we must have $\left(\delta^{2} M\right)_{S,C,z^i} \geq 0$, i.e.,
\begin{equation}
\left(\frac{\partial T}{\partial S}\right)_{C,z^i}^{(1)}+\left(\frac{\partial T}{\partial S}\right)_{C,z^i}^{(2)} \geq 0.
\end{equation}
Since the ordering of the subsystems is arbitrary, the above equation implies that $\left(\partial_S T\right)_{C,z^i} \geq 0$. At the stability limit, the equality holds:
\begin{equation}
\left(\frac{\partial T}{\partial S}\right)_{C,z^i} = 0, \label{eq:condtion_1}
\end{equation}
i.e. $\left(\delta^{2} M\right)_{S,C,z^i}=0$, and it further implies that
\begin{equation}
\left(\delta^{3} M\right)_{S,C,z^i}=\left(\delta S^{(1)}\right)^{3}\left[\left(\frac{\partial^{2} T}{\partial S^{2}}\right)_{C,z^i}^{(1)}-\left(\frac{\partial^{2} T}{\partial S^{2}}\right)_{C,z^i}^{(2)}\right] \geq 0.
\end{equation}
Following the same argument for the first-order term, we have
\begin{equation}
\left(\frac{\partial^{2} T}{\partial S^{2}}\right)_{C,z^i}^{(1)}-\left(\frac{\partial^{2} T}{\partial S^{2}}\right)_{C,z^i}^{(2)} = 0.
\end{equation}
Defining $s \equiv S/V_C$ and substituting this into the above equation, we have
\begin{equation}
\left(\frac{\partial^{2} T}{\partial s^{2}}\right)_{C,z^i}^{(1)}\left(\frac{1}{V^{(1)}_C}\right)^{2}-\left(\frac{\partial^{2} T}{\partial s^{2}}\right)_{C,z^i}^{(2)}\left(\frac{1}{V^{(2)}_C}\right)^{2}=0.
\end{equation}
where $V^{(1)}_C$ and $V^{(2)}_C$ refer to the thermodynamic volume in subsystems $1$ and $2$. Since $T^{(1)}=T^{(2)}=T$, $s^{(1)}=s^{(2)}=s$ at thermal equilibrium, and the division into subsystems is entirely arbitrary, we must have that
\begin{equation}
\left(\frac{\partial^{2} T}{\partial s^{2}}\right)_{C,z^i} =0= \left(\frac{\partial^{2} T}{\partial S^{2}}\right)_{C,z^i}.
\end{equation}
Until now, we have find that $\left(\delta M\right)_{S,C,z^i}=\left(\delta^{2} M\right)_{S,C,z^i}=\left(\delta^{3} M\right)_{S,C,z^i}=0$ at stability limit. To investigate the stability, one has to look at the fourth-order term:
\begin{equation}
\left(\delta^{2} M\right)_{S,C,z^i}=\left(\delta S^{(1)}\right)^{4}\left[\left(\frac{\partial^3 T}{\partial S^3}\right)_{C,z^i}^{(1)}+\left(\frac{\partial^3 T}{\partial S^3}\right)_{C,z^i}^{(2)}\right].
\end{equation}
Following the same argument for the second-order term, we must have $(\partial^3_S T)_{C,z^i} \geq 0$. In summary, for a stable black hole system,
\begin{equation}
\left(\frac{\partial T}{\partial S}\right)_{C,z^i}=0, \quad \left(\frac{\partial^2 T}{\partial S^2}\right)_{C,z^i}=0, \quad \left(\frac{\partial^3 T}{\partial S^3}\right)_{C,z^i} \geq 0, \label{eq:stable_con}
\end{equation}
need to be satisfied at the stability limit. If $(\partial^3_S T)_{C,z^i} > 0$, the system at this point is strictly stable, which corresponds to a strictly stable critical point. However, if $(\partial^3_S T)_{C,z^i} = 0$, we have to repeat a fluctuation analysis for high-order therms to justify the stability. Note that since $(\partial_S T)_{C,z^i}=(\partial_{r_+} S)^{-1}(\partial_{r_+} T)_{C,z^i}$ and $\partial_{r_+} S>0$, Eq. (\ref{eq:stable_con}) is equal to
\begin{equation}
\left(\frac{\partial T}{\partial r_+}\right)_{C,z^i}=0, \quad \left(\frac{\partial^2 T}{\partial r^2_{+}}\right)_{C,z^i}=0, \quad \left(\frac{\partial^3 T}{\partial r^3_{+}}\right)_{C,z^i} \geq 0.
\end{equation}

\section{Standard VdW like critical points} \label{VdW_point}

%%%%%%%%%%%%%%%%%%%%
\begin{figure}
\centering
\includegraphics[height=5cm]{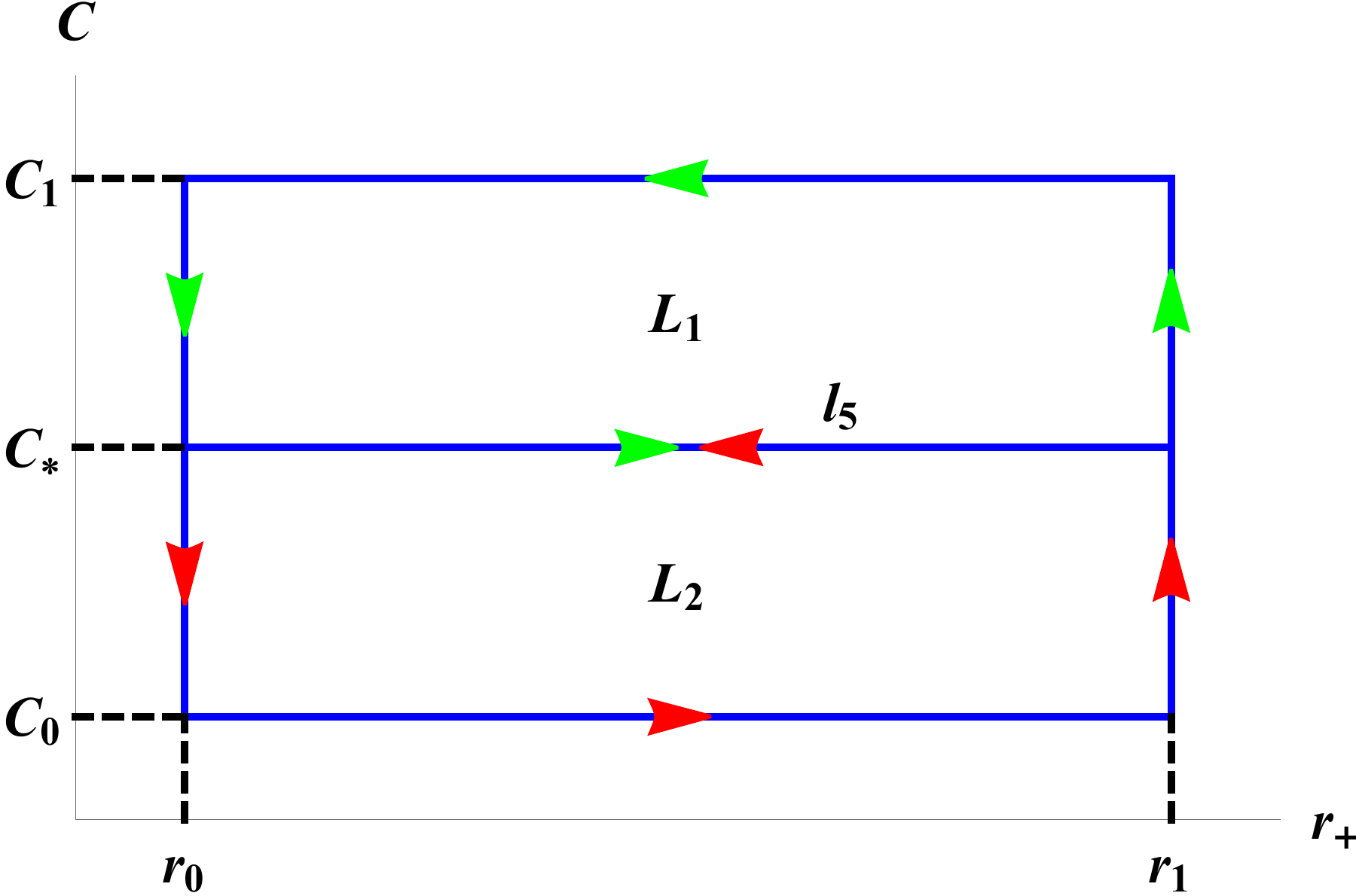}
\caption{\label{contour_L1_L2}Sketch map of the contours $L_1$ and $L_2$ on the $(r_+, C)$ plane. Green and red arrows refer to the positive orientation along $L_1$ and $L_2$, respectively.}
\end{figure}
%%%%%%%%%%%%%%%%%%%%%%%%

As discussed before, the standard VdW like critical points can appear when critical central charge $C<C_*=3 b^2 Q^2/2 \pi P$. To investigate the corresponding parameter range for $(Q,P,b)$, we can calculate the topological charge for the $C<C_*$ region in the $r_+-C$ plane, i.e. the $L_2$ region shown in Fig. \ref{contour_L1_L2}. If a positive topological charge is obtained, we can justify that at least one standard VdW like critical point exist.

Along $l_5=\{C=C_*, r_0=\delta \leq r_+ \leq r_1=1/\delta\}$ in Fig. \ref{contour_L1_L2}, we have
\begin{align}
&\Phi_1=\frac{1}{4 \pi} \left(2 b^2 + \frac{4 \pi  P}{b Q}-\frac{x+1}{r_+^2} +\frac{2}{x r_+^2}\right), \quad \Phi_2=\frac{\left(x-1\right) + 4 \left(x-3\right) b^4 r_+^4}{2 \pi x^{3} r_+^3},
\end{align}
where $x=\sqrt{1+4 b^4 r_+^4}$.
At the left end point of $l_5$, $r_+=\delta$, $C=C_*$, and
\begin{align}
\Phi_1(\delta \rightarrow 0) \sim \frac{b^2}{2 \pi} + \frac{P}{b Q}, \quad \Phi_2(\delta \rightarrow 0) \sim -\frac{3 b^4 }{\pi } \delta.
\end{align}
At the right end point of $l_5$, $r_+=1/\delta$, $C=C_*$, and
\begin{eqnarray}
\Phi_1(\delta \rightarrow 0) \sim \frac{P}{b Q}, \quad \Phi_2(\delta \rightarrow 0) \sim \frac{\delta^3}{2 \pi}.
\end{eqnarray}
Since $\Phi_2/\Phi_1 \rightarrow 0$ and $\Phi_1>0$, the vector $\Phi$ is horizontal to the right at these points. This indicates that $\Omega$ may changes $2k \pi$ in clockwise direction or anti-clockwise direction, where $k \in (0,1,2,\cdots)$. To further obtain the rotating direction and the value of $k$, one has to study the the zeros of $\Phi_1$ and $\Phi_2$. For $\Phi_1$, due to the complexity, it is difficult to get analytical zeros. Fortunately, the equation $\Phi_2=0$ can be cast into a quadratic equation, and the solution reads
\begin{eqnarray}
r_{\text{zero}}=\left(\frac{2 \sqrt{3}+3}{4}\right)^{\frac{1}{4}} \frac{1}{b}.
\end{eqnarray}
Note that form our definition for $\Phi_1$ and $\Phi_2$, $\Phi_2$ can be identified the derivative of $\Phi_1$ respect to $r_+$, and thus $r_{\text{zero}}$ corresponds to the (unique) extremum of $\Phi_1$. Detailed analysis shows that $\Phi_2(r_+ < r_{\text{zero}})<0$ and $\Phi_2(r_+ > r_{\text{zero}})>0$, thus $r_{\text{zero}}$ exactly corresponds to the global minimum of $\Phi_1$. Substituting $r_{\text{zero}}$ into $\Phi_1$, we have
\begin{align}
&\Phi_1 (r_+=r_{\text{zero}})=\frac{P}{b Q} +\frac{\left[3 \left( \sqrt{2 \sqrt{3}+3}+ \sqrt{4-2 \sqrt{3}}-1 \right) -3^{3/4} \sqrt{4 \sqrt{3}+6}\right] b^2}{6 \pi \sqrt{2 \sqrt{3}+3} }.
\end{align}
If the above equation is larger than $0$, i.e., $P > 0.029 b^3 Q$, $\Phi_1$ does not possess any zero point (since $\Phi_1(r_+ \rightarrow 0^+/+\infty) >0 $), which indicates that the vector $\Phi$ can not wind beyond $2 \pi$, i.e., $k=0$. If the above equation is smaller than $0$, i.e., $P < 0.029 b^3 Q$, $\Phi_1$ possesses two zero points, indicates that the vector $\Phi$ winds $2 \pi$, i.e., $k=1$. From left to right, i.e. along the positive orientation on contour $L_1$, $\Phi$ winds in the clockwise direction, and on the opposite, along the positive orientation on contour $L_2$, $\Phi$ winds in the anti-clockwise direction. Combing the analysis on $l_1 \sim l_4$, we can find
\begin{align}
&Q_{L_1} = \left\{\begin{array}{ll}
0 & \text { for } P > 0.029 b^3 Q, \\
-1 & \text { for } P < 0.029 b^3 Q,
\end{array}\right.\\
&Q_{L_2} = \left\{\begin{array}{ll}
0 & \text { for } P > 0.029 b^3 Q, \\
+1 & \text { for } P < 0.029 b^3 Q.
\end{array}\right.
\end{align}
This suggests that when $P<0.029 b^3 Q$, there will be at least one standard VdW like critical point in the $C<C_*$ region.

\end{document}